\documentclass[aps,prx,showpacs,amssymb,amsart,superscriptaddress,nofootinbib,twocolumn,longbibliography,floatfix]{revtex4-1}

\usepackage{amsmath,amsfonts,amsthm,amssymb}
\usepackage{braket}
\usepackage{float}
\usepackage{setspace}
\usepackage{amsmath}
\usepackage{appendix}
\usepackage[ansinew]{inputenc}
\usepackage{bbm}
\usepackage{bm}
\usepackage{amsbsy}
\usepackage{dsfont} 
\usepackage{graphicx} 
\usepackage{epsfig}
\usepackage{epstopdf}
\usepackage{dsfont}
\usepackage{color}
\usepackage[bookmarks=true, colorlinks,breaklinks]{hyperref}
\makeatletter
\newcommand\org@hypertarget{}
\let\org@hypertarget\hypertarget
\renewcommand\hypertarget[2]{%
  \Hy@raisedlink{\org@hypertarget{#1}{}}#2%
  }
\makeatother

\usepackage[figure,table]{hypcap}
\usepackage{enumerate}
\usepackage[resetlabels]{multibib}

\hypersetup{
	bookmarksnumbered,
	pdfstartview={FitH},
	citecolor={darkgreen},
	linkcolor={darkred},
	urlcolor={darkblue},
	pdfpagemode={UseOutlines}}
\definecolor{darkgreen}{RGB}{50,190,50}
\definecolor{darkblue}{RGB}{0,0,190}
\definecolor{darkred}{RGB}{238,0,0}
\definecolor{mycolor}{rgb}{0.122, 0.435, 0.698}
\definecolor{noeyecancergreen}{RGB}{93,240,88}
\usepackage{soul}

\newcommand{\scpr}[2]{\ensuremath{\left\langle\right.\! #1 \!\left|\right.\! #2 \!\left.\right\rangle}}

\newcommand{\expval}[1]{\ensuremath{\left\langle\right.\! #1 \!\left.\right\rangle}}

\newcommand{\pr}{^{\prime}}

\newcommand{\subtiny}[3]{\ensuremath{_{\hspace{#1 pt}\protect\raisebox{#2 pt}{\tiny{$ #3$}}}}}

\newcommand{\tr}{\textnormal{Tr}}
\newcommand{\djj}{d\kern-0.4em\char"16\kern-0.1em}
\usepackage{tabularx}
\usepackage{tcolorbox}

\usepackage{paracol}
\tcbuselibrary{theorems}

\newtcbtheorem{Boxes}{Box}%
{colback=mycolor!5,colframe=mycolor,fonttitle=\bfseries,
left=4pt,
right=4pt,
top=5pt,
bottom=5pt,float=ht}{boxbox}

\newcommand{\iqoqi}{Institute for Quantum Optics and Quantum Information (IQOQI), Austrian Academy of Sciences, Boltzmanngasse 3, A-1090 Vienna, Austria}
\newcommand{\hwu}{Institute of Photonics and Quantum Sciences (IPaQS), Heriot-Watt University, Edinburgh, EH14 4AS, UK}

\usepackage[framemethod=tikz]{mdframed}
\definecolor{mycolor}{rgb}{0.122, 0.435, 0.698}
\newmdenv[innerlinewidth=0.5pt, roundcorner=4pt,linecolor=mycolor,innerleftmargin=6pt,
innerrightmargin=6pt,innertopmargin=6pt,innerbottommargin=6pt]{mybox}
\usepackage{paracol}
\usepackage{tcolorbox}
\tcbuselibrary{theorems}

\newtcbtheorem{Definitions}{Definition}%
{colback=mycolor!5,colframe=mycolor,fonttitle=\bfseries,
left=4pt,
right=4pt,
top=5pt,
bottom=5pt}{defi}

\newtcbtheorem{Lemmas}{Lemma}%
{colback=green!5,colframe=green!50!blue,fonttitle=\bfseries,
left=4pt,
right=4pt,
top=5pt,
bottom=5pt
}{lemma}

\newtcbtheorem{BBoxes}{Box}%
{colback=mycolor!5,colframe=mycolor,fonttitle=\bfseries,
left=4pt,
right=4pt,
top=5pt,
bottom=5pt,float=ht!}{boxbox}

\newtcbtheorem{WideBoxes}{Box}%
{width=\textwidth,colback=mycolor!5,colframe=mycolor,fonttitle=\bfseries,
left=4pt,
right=4pt,
top=5pt,
bottom=5pt,float*=ht!}{boxbox}

\newtcolorbox[blend into=figures]{boxdefi}[3][]
{ float*=ht,width=\textwidth,lower separated=false, center upper,
title={#2},label= def:#3,#1}

\newtcolorbox[blend into=tables]{smallboxtable}[3][]
{colback=mycolor!5, colframe=mycolor, float*=ht, width=\textwidth, lower separated=false, blend before title=colon hang,
title={#2}, label= table:#3 ,#1}

\newcolumntype{Y}{>{\centering\arraybackslash}X}

\begin{document}

%
\title{Entanglement Certification -- From Theory to Experiment}
\author{Nicolai Friis}
\email{nicolai.friis@univie.ac.at}
\affiliation{\iqoqi}
\author{Giuseppe Vitagliano}
\email{giuseppe.vitagliano@univie.ac.at}
\affiliation{\iqoqi}
\author{Mehul Malik}
\email{m.malik@hw.ac.uk}
\affiliation{\iqoqi}
\affiliation{\hwu}
\author{Marcus Huber}
\email{marcus.huber@univie.ac.at}
\affiliation{\iqoqi}
\maketitle

\textbf{Entanglement is an important resource that allows quantum technologies to go beyond the classically possible. There are many ways quantum systems can be entangled, ranging from the archetypal two-qubit case to more exotic scenarios of entanglement in high dimensions or between many parties. Consequently, a plethora of entanglement quantifiers and classifiers exist, corresponding to different operational paradigms and mathematical techniques. However, for most quantum systems, exactly quantifying the amount of entanglement is extremely demanding, if at all possible. This is further exacerbated by the difficulty of experimentally controlling and measuring complex quantum states. Consequently, there are various approaches for experimentally detecting and certifying entanglement when exact quantification is not an option, with a particular focus on practically implementable methods and resource efficiency. The applicability and performance of these methods strongly depends on the assumptions one is willing to make regarding the involved quantum states and measurements, in short, on the available prior information about the quantum system. In this review we discuss the most commonly used paradigmatic quantifiers of entanglement. For these, we survey state-of-the-art detection and certification methods, including their respective underlying assumptions, from both a theoretical and experimental point of view.}


In the early twentieth century, the phenomenon of quantum entanglement rose to prominence as a central feature of the famous thought experiment by Einstein, Podolsky, and Rosen~\cite{EinsteinPodolskyRosen1935}. Initially disregarded as a mathematical artefact that showcases the incompleteness of quantum theory, the properties of entanglement were largely ignored until 1964, when John Bell famously proposed an experimentally testable inequality able to distinguish between the predictions of quantum mechanics and those of any local-realistic theory~\cite{Bell1964}. With the advent of the first experimental tests~\cite{FreedmanClauser1972}, spearheaded by Aspect \textit{et al.}~\cite{AspectGrangierRoger1981, AspectGrangierRoger1982, AspectDalibardRoger1982},
emerged the realisation that entanglement constitutes a resource for information processing and communication tasks~\cite{Bennettetal1993, BennettBrassard1984, BennettBrassard2014QuantumCP, Ekert1991}, which was in turn confirmed empirically in a series of ground-breaking experiments~\cite{Bouwmeester-etal1997, WeihsJenneweinSimonWeinfurterZeilinger1998, PoppeEtAl2004, Ursin-Zeilinger2007}. Since this advent of quantum information theory~\cite{NielsenChuang2000}, the field has advanced, diversified, and many links have been established with other disciplines. Today, the study of Bell-like inequalities is an active field of research~\cite{BrunnerCavalcantiPironioScaraniWehner2014}, culminating in recent experimental tests closing all loopholes~\cite{Hensen-Hanson2015, Giustina-Zeilinger2015, Shalm-Nam2015}. Entanglement was thus once and for all proven to be an indispensable ingredient for the description of nature. Moreover, it is now clear that modern quantum technologies have the capability of producing, manipulating, and certifying entanglement.

In the early days of quantum information, Werner realised that entanglement and the violation of Bell inequalities are not necessarily the same phenomenon~\cite{Werner1989}. While entanglement is needed to violate Bell inequalities, it is still not known if (and in what sense) entanglement always allows for Bell violation~\cite{Barrett2002, AcinGisinToner2006, Vertesi2008, HirschQuintinoVertesiNavascuesBrunner2016}. From a contemporary perspective, Bell inequalities are seen as device-independent certifications of entanglement. The question whether all entangled states can be certified device-independently is hence still an open problem. In his seminal paper~\cite{Werner1989}, Werner also gave the first formal mathematical definition of entanglement. Since then, entanglement theory as a means to characterise and quantify entanglement has developed into an entire sub-field of quantum information. Previous reviews have captured various aspects of the research in this sub-field, focusing, for example, on the nature of non-entangled states~\cite{Bruss2002}, the quantification of entanglement as a resource~\cite{PlenioVirmani2007, EltschkaSiewert2014}, or providing detailed collections of works on entanglement theory~\cite{HorodeckiEntanglementReview2009} and entanglement detection~\cite{GuehneToth2009}.

In this review we address the current challenges of experimentally certifying and quantifying entanglement in quantum systems too complex for conventional tomography to be a feasible option. These challenges already arise for finite-dimensional systems, which we focus on here, referring the interested reader to existing reviews on continuous-variable entanglement~\cite{BraunsteinVanLoock2005, FerraroOlivaresParis2005, AdessoIlluminati2007, AndersenLeuchsSilberhorn2010, WeedbrookPirandolaGarciaPatronCerfRalphShapiroLloyd2012, AdessoRagyLee2014}, for a discussion of the fascinating intricacies of infinite-dimensional systems. With the advent of the first large scale quantum devices and the increased complexity of manufactured quantum technologies, this is a field of growing importance. In quantum communication, certifiable entanglement forms the basis for the next generation of secure quantum devices~\cite{CerfBourennaneKarlssonGisin2002, BarrettKentPironio2006, GroeblacherJenneweinVaziriWeihsZeilinger2006, HuberPawlowski2013}. Here, it is important to note that entanglement certification goes beyond entanglement estimation in the sense that the latter may rely on reasonable assumptions about the system state or measurement setup, whereas requirements for certification are stricter. In quantum computation, the certified presence of entanglement points towards the use of actual quantum resources, which is crucial if one is to trust the devices' correct functionality~\cite{JozsaLinden2003}, while in quantum simulation~\cite{Lloyd1996, Lanyon-Roos2011, Barreiro-Blatt2011, CiracZoller2012, Britton-Bollinger2012} a large amount of entanglement can serve as an indicator of the hardness of classically simulating the corresponding states~\cite{AmicoFazioOsterlohVedral2008,VerstraeteMurgCirac2008,EisertCramerPlenio2010,Schollwoeck2011,Orus2014,RanTirritoPengChenSuLewenstein2017}. Nonetheless, the precise role of entanglement in quantum computation and simulation is less clearly delineated as it is in quantum communication. Finally, entanglement can be understood as a means of bringing about speed-ups~\cite{WinelandBollingerItanoMooreHeinzen1992,  HuelgaMacchiavelloPellizzariEkertPlenioCirac1997, GiovannettiLloydMaccone2006}, parallelisation~\cite{Maccone2013}, and even flexibility~\cite{FriisOrsucciSkotiniotisSekatskiDunjkoBriegelDuer2017} in quantum metrology~\cite{TothApellaniz2014}. Non-coincidentally, these four areas also form the central pillars of the recent European flagship program on quantum technologies~\cite{FlagshipRoadmap2018}.


\section*{Detection \& quantification of entanglement}

\textbf{Entanglement \& separability}.\ Entanglement is conventionally defined via its contrapositive: \emph{separability}. A pure quantum state is called separable with respect to a tensor factorisation $\mathcal{H}\subtiny{0}{0}{A}\otimes\mathcal{H}\subtiny{0}{0}{B}$ of its (finite-dimensional) Hilbert space if and only if it can be written as a product state  $\ket{\psi}\subtiny{0}{0}{AB}:=\ket{\phi}\subtiny{0}{0}{A}\otimes\ket{\chi}\subtiny{0}{0}{B}$. A general (mixed) quantum state $\rho$ is called separable if it can be written as a probabilistic mixture of separable pure states~\cite{Werner1989}
\begin{align}
    \rho_{\mathrm{sep}}:=\sum_i p_i
    \ket{\phi_{i}}\!\!\bra{\phi_{i}}\subtiny{0}{0}{A}
    \otimes
    \ket{\chi_{i}}\!\!\bra{\chi_{i}}\subtiny{0}{0}{B},
    \label{eq:sep}
\end{align}
All of the infinitely many pure state decompositions of a density matrix can be interpreted as a concrete instruction for preparing the quantum state via mixing the states $\ket{\phi_{i}}\subtiny{0}{0}{A}\ket{\chi_{i}}\subtiny{0}{0}{B}$ drawn from a classical probability distribution $\{p_{i}\}_{i}$. Since each of these pure states is separable, mixed separable states can easily be prepared by coordinated local operations, i.e., local operations aided by classical communication (LOCC)~\cite{Nielsen1999, ChitambarLeungMancinskaOzolsWinter2014}. Conversely, any state that is not separable is called \emph{entangled} and can not be created via LOCC. The fact that there are infinitely many ways to decompose a density matrix into pure states is at the root of the central challenge in entanglement theory: To conclude that a state is indeed entangled one needs to rule out that \textemdash\ among infinitely many \textemdash\ there is \emph{any} decomposition into product states. Answering this question for general density matrices is an NP-hard problem~\cite{Gurvits2003, Gurvits2004}. To be precise, even the relaxed problem allowing for a margin of error that is inversely polynomial (in contrast to inverse exponential errors in the original proof by Gurvits) in the system dimension remains NP-hard~\cite{Gharibian2010}.

Pure states, separable or entangled, admit a Schmidt decomposition into bi-orthogonal product vectors, i.e., we can write them as $\ket{\psi}\subtiny{0}{0}{AB}=\sum_{i=0}^{k-1}\lambda_i\ket{ii}$. The coefficients $\lambda_i\in\mathbb{R}^{+}$ are called Schmidt coefficients. Their squares, which are equal to the eigenvalues of the marginals $\rho\subtiny{0}{0}{A/B}:=\text{Tr}\subtiny{0}{0}{B/A}\ket{\psi}\!\!\bra{\psi}\subtiny{0}{0}{AB}$, are usually arranged in decreasing order and collected in a vector $\vec{\lambda}$ with components $[\vec{\lambda}]_i:=\lambda_i^2$. The number $k$ of non-zero Schmidt coefficients is called \emph{Schmidt rank}, or sometimes dimensionality of entanglement, as it represents the minimum local Hilbert space dimension required to faithfully represent the correlations of the quantum state. One of the fundamental pillars of state manipulation under LOCC is Nielsen's majorisation theorem~\cite{Nielsen1999, NielsenVidal2001}: \textit{A quantum state with Schmidt coefficients $\{\vec{\lambda}_i\}_{i}$ can be transformed to another state with Schmidt coefficients $\{\vec{\lambda\pr}_j\}_{j}$ via an LOCC transformation if and only if $\vec{\lambda}\prec\vec{\lambda\pr}$, i.e., the vector of squared Schmidt coefficients of the output state majorises the corresponding vector of the input state}. This also conveniently captures two extremal cases: A separable state has a corresponding vector of $(1,0,...,0,)$, majorising every other vector, and thus cannot be transformed into any entangled state via LOCC. In dimensions $d$, the vector $(\tfrac{1}{d},\tfrac{1}{d},...,\tfrac{1}{d})$ on the other hand is majorised by every other vector. The corresponding state $\ket{\Phi^{+}}:=\tfrac{1}{\sqrt{d}}\sum_{i=0}^{d-1}\ket{ii}$ can thus be transformed into \emph{any} other quantum state and is therefore referred to as a \emph{maximally entangled state}.


\textbf{Entanglement quantification}.\ Any meaningful entanglement quantifier for pure states is hence a function of the Schmidt coefficients. The two most prominent representative are the entropy of entanglement, i.e., the von~Neumann entropy of the marginals, or equivalently the Shannon entropy of the squared Schmidt coefficients $E(\ket{\psi}\subtiny{0}{0}{AB}):=S(\rho\subtiny{0}{0}{A/B})=-\sum_{i=0}^{k-1}\lambda_{i}^{2}\log_{2}(\lambda_{i}^{2})$, and the R{\' e}nyi $0$-entropy or the logarithm of the marginal rank. For mixed states, the fact that there exist infinitely many pure state decompositions complicates the quantification of entanglement. How is one to unambiguosly quantify the entanglement of a state that admits different decompositions into states with various degrees of entanglement? A straightforward answer presents itself in the form of an average over the entanglement $E(\ket{\psi_{i}})$ within a given decomposition, minimised over all decompositions $\mathcal{D}(\rho)$, i.e.,  $E(\rho):=\inf_{\mathcal{D}(\rho)}\sum_{i}p_{i} E(\ket{\psi_{i}})$. When choosing the entropy of entanglement as the measure of choice, this convex roof construction leads to the \emph{entanglement of formation} $E_{\mathrm{oF}}$~\cite{BennettDiVincenzoSmolinWootters1996, Wootters1998}. Its regularisation $\lim_{n\rightarrow\infty}\tfrac{1}{n}E_{\mathrm{oF}}(\rho^{\otimes n})$ has a convenient operational interpretation as the \emph{entanglement cost}~\cite{BennettDiVincenzoSmolinWootters1996, HaydenHorodeckiTerhal2001}, i.e., the asymptotic LOCC interconversion rate from $m$ qubit Bell states $\ket{\psi}^{\otimes m}=\tfrac{1}{\sqrt{2}}(\ket{00}+\ket{11})^{\otimes m}$ to $n$ copies of $\rho$, i.e., $\rho^{\otimes n}$. Conversely, one may define distillable entanglement as the asymptotic LOCC conversion rate from nonmaximally entangled states to Bell states~\cite{BennettBernsteinPopescuSchumacher1996, BennettBrassardPopescuSchumacherSmolinWooters1996}. If the $E_{\mathrm{oF}}$ were additive, it would coincide with the entanglement cost. However, as shown by Hastings~\cite{Hastings2009}, the entanglement of formation is only sub-additive. For other measures, such as the Schmidt rank, a more appropriate generalisation is to maximise (instead of averaging) over all states within a given decomposition. In this way, the \emph{Schmidt number} of mixed quantum states, defined as $d_{\rm{ent}}:=\inf_{\mathcal{D}(\rho)}\max_{\ket{\psi_{i}}\in\mathcal{D}(\rho)}\operatorname{rank}\bigl(\tr_{A}(\ket{\psi_{i}}\!\!\bra{\psi_{i}})\bigr)$~\cite{TerhalHorodecki2000}, directly inherits the operational interpretation of the Schmidt rank for pure states. These are just two exemplary cases out of a plethora of generally inequivalent entanglement measures and monotones. For an in-depth review, we refer the interested reader to~\cite{PlenioVirmani2007, EltschkaSiewert2014}. While these and many other measures have very instructive and operational interpretations, even deciding whether they are non-zero is an NP-hard problem in general, even if the density matrix is known to infinite precision. However, not only will uncertainties be associated to the different matrix elements obtained in actual experiments, the sheer amount of information that needs to be collected renders full state tomography too cumbersome to be practical beyond small-scale demonstrations~\cite{AltepeterJamesKwiat2004, Ansmann-Martinis2006}. This is exacerbated in the multipartite case, where the system dimension grows exponentially with the number of parties.

An implication of this observation is that the amount of actual entanglement in a quantum system not only depends on the measure used (and hence the context or task for which it is applied), but is also impossible to ascertain exactly. However, it is possible to certify the presence and even lower-bound the amount of entanglement for various useful quantifiers through few experimentally realisable measurements, which will be the main focus of this review.


\textbf{Partial transposition \& entanglement distillation}.\ A recurring feature among entanglement tests is to overcome the hardness of the separability problem by detecting only a subset of entangled states. An example (that nonetheless requires knowledge of the entire density matrix) is the \emph{positive partial transpose} (PPT) criterion~\cite{Peres1996, HorodeckiMPR1996}. That is, partially transposing a separable state leads to a positive semi-definite density matrix, but this need not be the case for entangled states. This is because the partial transposition is an instance of a positive, but not completely positive map. One the one hand, positive maps $\Lambda_{\mathrm{P}}[\rho]\geq0$ lead to positive semi-definite matrices when applied to positive semi-definite matrices, such as quantum states. Completely positive maps $(\Lambda_{\mathrm{CP}}\otimes\mathbbm{1}_d)[\rho]\geq0$ $\forall d\in \mathbb{Z}^+$, on the other hand, lead to positive semi-definite operators even when applied to marginals. In fact, it was proven that a state is separable if and only if it remains positive under \emph{all} positive maps applied to a subsystem~\cite{HorodeckiMPR1996}.

In addition to serving as an easily implementable entanglement test (provided the density matrix is known), the partial transposition provides a simple sufficient criterion for distillation. As shown in Ref.~\cite{HorodeckiMPR1998}, the process of \emph{entanglement distillation}~\cite{BennettBernsteinPopescuSchumacher1996, BennettBrassardPopescuSchumacherSmolinWooters1996}, i.e., the simultaneous local processing of multiple copies of pairwise distributed quantum states to concentrate the entanglement in one pair, is only possible if there exists at least a $2\times2$-dimensional subspace of the multi-copy state space that is not PPT. Since any tensor products of PPT states are also PPT, this directly implies that even though many PPT states are entangled, none of them are distillable. Conversely, whether all states which are non-positive under partial transposition (NPT) are distillable is still an open problem~\cite{PankowskiPianiHorodeckiMP2010}, but it is known that for any finite number of copies the answer is negative~\cite{Watrous2004}.

The PPT map is also commonly used to quantify entanglement via the logarithmic negativity~\cite{Plenio2005}, defined as the logarithm of the trace norm of the partially transposed density matrix, i.e., $\mathcal{N}(\rho):=\log_2(||\Lambda_{\mathrm{P}}[\rho]||_1)$. Loosely speaking, it captures how much the partial transpose fails to be non-negative. The logarithmic negativity is a prominent example of an \emph{entanglement monotone}~\cite{Vidal2000} (as is the negativity~\cite{EisertPhD2001, VidalWerner2002}), i.e., a quantity that is non-increasing under LOCC like any entanglement measure, but which need not necessarily be non-zero for all entangled states.

Whereas calculating the result of applying a positive map requires knowledge of the entire density matrix, it is still possible to harness positive maps to construct powerful entanglement witnesses~\cite{HorodeckiMPR1996} even if only partial or imprecise information about the state is available. Suppose one is provided with a theoretical target state $\rho\subtiny{0}{0}{\mathrm{T}}$ that is not positive semi-definite under a positive (but not completely positive) map $\Lambda_{\mathrm{P}}$, i.e., $\Lambda_{\mathrm{P}}[\rho\subtiny{0}{0}{\mathrm{T}}]\ngeq0$. Then there exist vectors (e.g., preferably the eigenvector $\ket{\psi^{-}}$ of $\Lambda_{\mathrm{P}}[\rho\subtiny{0}{0}{\mathrm{T}}]$ corresponding to the smallest eigenvalue) for which $\bra{\psi^{-}} \Lambda_{\mathrm{P}}[\rho\subtiny{0}{0}{\mathrm{T}}] \ket{\psi^{-}} = \tr\bigl(\Lambda_{\mathrm{P}}[\rho\subtiny{0}{0}{\mathrm{T}}] \ket{\psi^{-}}\!\!\bra{\psi^{-}}\bigr)<0$. Via the dual map $\Lambda_{\mathrm{P}}^{*}$ this is equivalent to the statement $\tr\bigl(\rho\subtiny{0}{0}{\mathrm{T}} \Lambda_{\mathrm{P}}^{*}[\ket{\psi^{-}}\!\!\bra{\psi^{-}}]\bigr)<0$, whereas
$\tr\bigl(\sigma \Lambda_{\mathrm{P}}^{*}[\ket{\psi^{-}}\!\!\bra{\psi^{-}}]\bigr)\geq0$ for all separable states $\sigma$. The Hermitian operator $\Lambda_{\mathrm{P}}^{*}[\ket{\psi^{-}}\!\!\bra{\psi^{-}}]$ is thus an example for an entanglement witness (see Box~\ref{boxbox:witnesses}), i.e., an observable that can in principle be measured to detect entangled states at least in the vicinity of $\rho\subtiny{0}{0}{\mathrm{T}}$.


\begin{WideBoxes}{\rm Entanglement Witnesses}{witnesses}
\emph{Entanglement witnesses}~\cite{Terhal2000} constitute one of the most important practical entanglement certification techniques. The key idea behind this concept is of geometrical nature. The definition of separability in Eq.~(\ref{eq:sep}) implies that the set $\mathcal{S}$ of separable states is a convex subset of all quantum states. The Hahn-Banach theorem~\cite[pp.~75]{ReedSimon1972} then guarantees that there exists a hyperplane for every entangled state $\rho$ that separates this state from the separable set. These hyperplanes correspond to observables $W$, such that $\tr(W\rho)<0$, whereas $\tr(W\sigma)\geq0$ for all $\sigma\in\mathcal{S}$. Measurements of such witness operators can hence certify the presence of entanglement.

That is, having identified an operator $W$ whose expected value is non-negative for all separable states, measuring $\expval{W}_{\rho}$ and obtaining a negative value conclusively demonstrates that $\rho$ is entangled. Although such witnesses exist for every entangled state $\rho$, finding $\expval{W}_{\rho}\geq0$ does not imply that $\rho$ is separable: $W$ might simply not be a suitable witness for the underlying state. The challenge hence lies in the construction of useful entanglement witnesses. Without specific information about the state produced in an experiment, this is a formidable task. However, when the underlying state can be expected to be close to a target state $\ket{\psi\subtiny{0}{0}{\mathrm{T}}}$, there exists a canonical witness construction, given by
\begin{align}
    W:=\lambda_{\mathrm{max}}^{2}\mathbbm{1}-\ket{\psi\subtiny{0}{0}{\mathrm{T}}}\!\!\bra{\psi\subtiny{0}{0}{\mathrm{T}}}.
\end{align}
Here, $\lambda_{\mathrm{max}}$ is the largest Schmidt coefficient of $\ket{\psi\subtiny{0}{0}{\mathrm{T}}}$, representing the maximal overlap of any separable state with $\ket{\psi\subtiny{0}{0}{\mathrm{T}}}$, such that $\max_{\sigma\in\mathcal{S}}\tr(\sigma W)=0$. While entanglement witnesses are observables and can hence in principle be evaluated by measurements in only a single basis, the corresponding basis cannot be a product basis, but must consist (at least in part) of basis states featuring entanglement across the partition for which entanglement is to be detected in the first place. More specifically, we can express any witness $W$ for entanglement across a bipartition $A|B$ w.r.t. local operator bases $\{g_{A}^{i}\}_{i}$ and $\{g_{B}^{j}\}_{j}$ (e.g., appropriately normalised Pauli matrices for qubits) with $\tr(g_{A/B}^{i}g_{A/B}^{j})=d\delta_{ij}$, that is $W=\sum_{i,j}c_{ij}g_{A}^{i}\otimes g_{B}^{j}$. This means that entanglement witnesses can also be obtained by a larger number of local measurements, where the figure of merit is the number of non-zero coefficients $c_{ij}$, determining the overall number of local measurement settings required to evaluate the witness (see Tab.~\ref{table:ms}). Bell-type inequalities are examples for such locally measured witnesses, e.g., the Clauser-Horne-Shimony-Holt inequality~\cite{ClauserHorneShimonyHolt1969} requires the measurement of $2$ product observables. Entanglement witnesses thus connect practical detection with geometric~\cite{BertlmannNarnhoferThirring2002, FriisBulusuBertlmann2017} and foundational aspects of entanglement.
    \begin{center}
    (a)\includegraphics[width=0.35\textwidth,trim={0cm 1cm 0cm 0.1cm},clip]{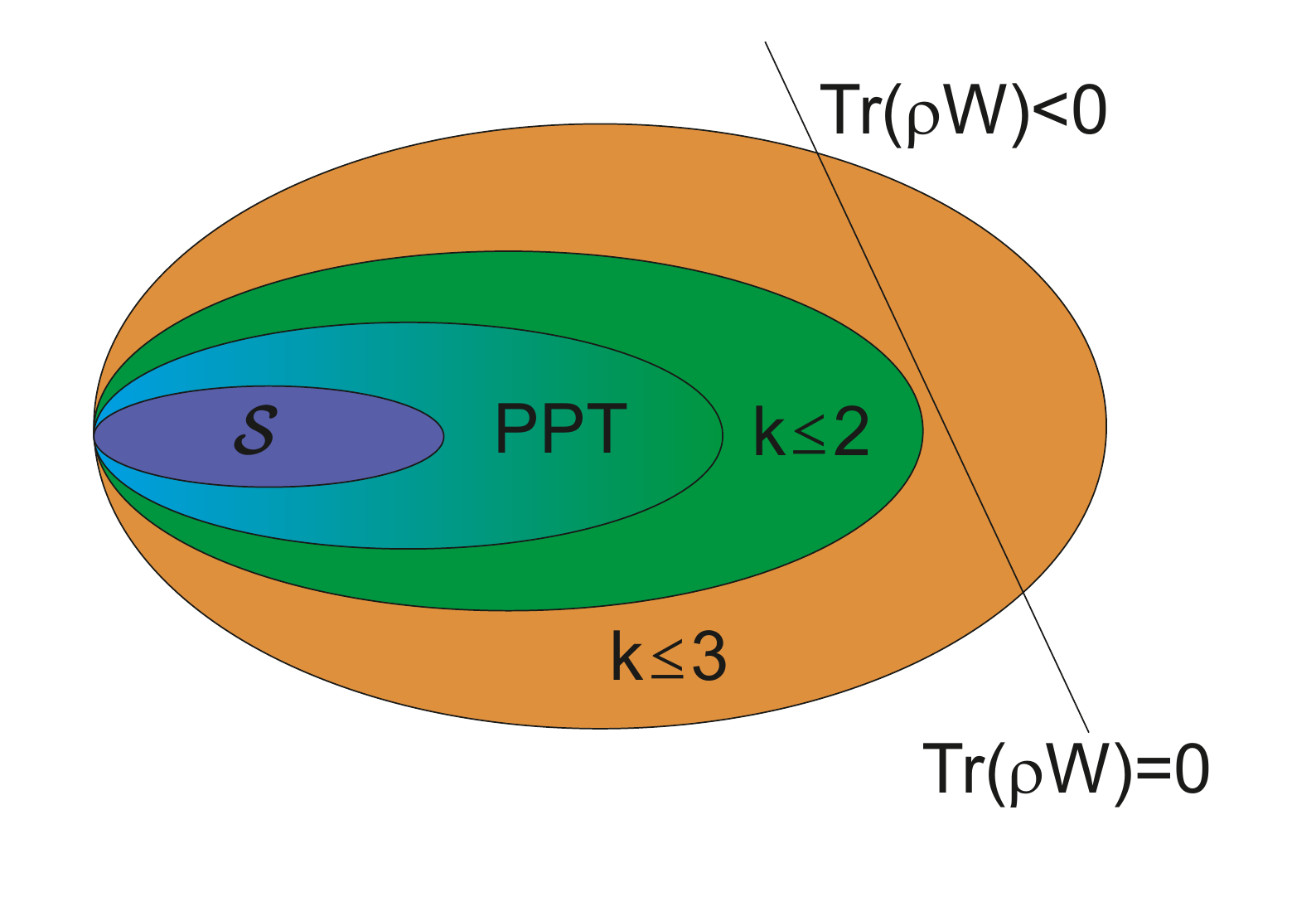}
    (b)\includegraphics[width=0.35\textwidth,trim={0cm 1cm 0cm 0.1cm},clip]{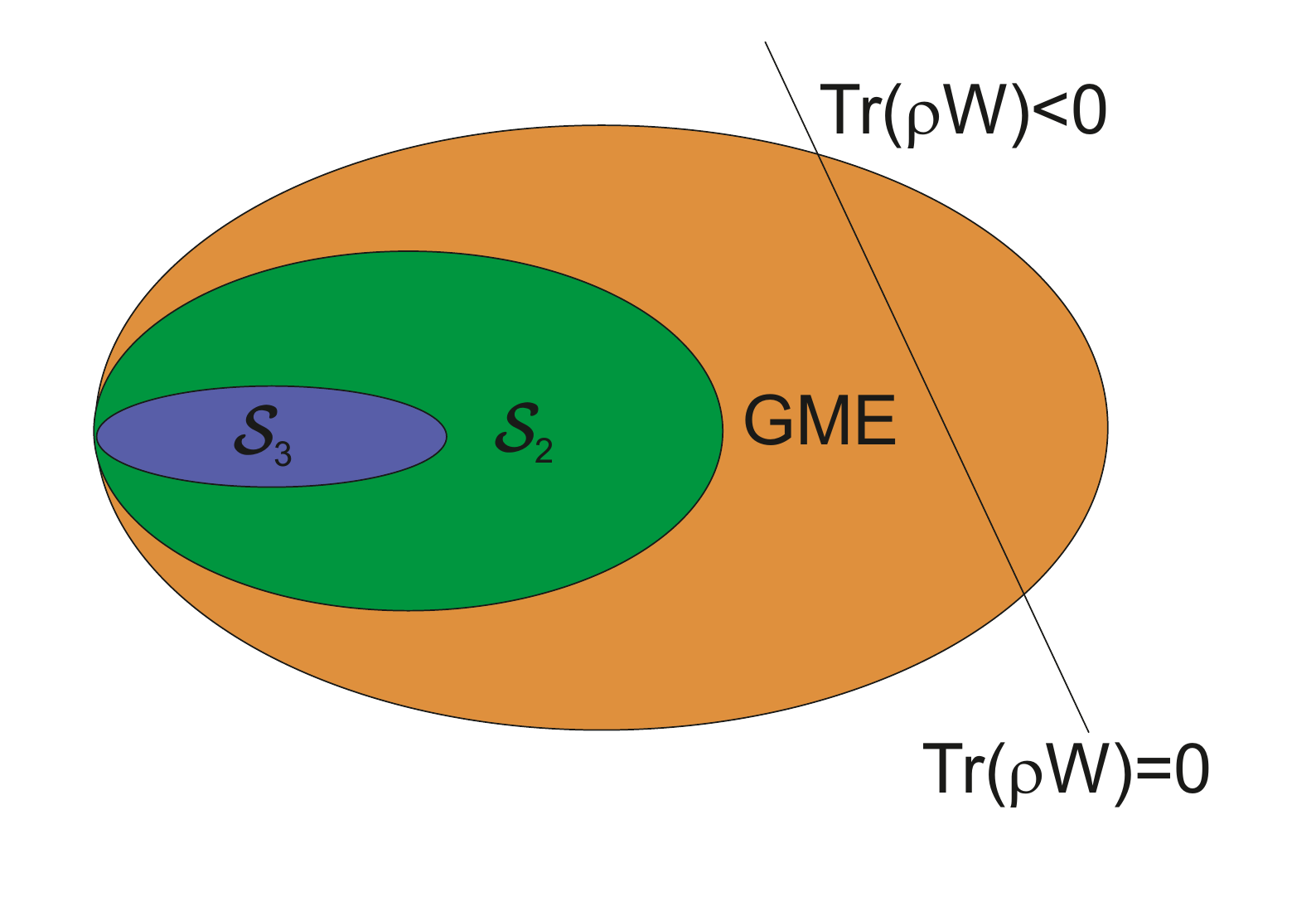}
    \end{center}
    \small{The illustration in (a) shows the nested convex structure of a $3\times3$-dimensional Hilbert space, i.e., the set of separable states $\mathcal{S}$ (with Schmidt number $k=1$, blue), the set of states positive under partial transposition (PPT, containing $\mathcal{S}$), the set of states with Schmidt number $k\leq2$ (green, containing PPT entangled states for which $k=2$), and the set of states with Schmidt number $k\leq3$ (orange, containing all others). The entanglement witness $W$ shown is an example for a Schmidt number witness, certifying genuine three-dimensional entanglement. Illustration (b) shows the nested convex structure of multipartite entanglement in a three-party Hilbert space, showing the set of fully separable states ($\mathcal{S}_3$, blue), biseparable states ($\mathcal{S}_2$, green) and genuine multipartite entanglement (GME, orange). The entanglement witness $W$ shown is an example for a multipartite entanglement witness certifying genuine three-partite entanglement.}
\end{WideBoxes}


\textbf{Linear contractions \& realignment}.\ Complementary to the partial transpose and other positive but not completely positive maps, there is another class of linear maps that provide a useful (and also complete) set of entanglement criteria: the {\it linear contractions}, i.e., linear maps that do not increase the trace norm of product states $\|\Lambda_{T} [\sigma\subtiny{0}{0}{A} \otimes \sigma\subtiny{0}{0}{B}]\|_1 \leq 1$. This property extends to all separable states, that is, it can be proven that a state $\sigma\subtiny{0}{0}{AB}$ is separable {\it if and only if} it satisfies $\|\Lambda_{T}[\sigma\subtiny{0}{0}{AB}]\|_1 \leq 1$ for all linear maps that do not increase the trace norm of product states. Consequently, such maps can be used to detect those states $\rho\subtiny{0}{0}{AB}$ for which $\|\Lambda_{T} [\rho\subtiny{0}{0}{AB}]\|_1 > 1$, which must hence be entangled~\cite{HorodeckiMPR2006}.

A prominent example of the above method is the so called {\it computable cross-norm or realignment} (CCNR) criterion~\cite{Rudolph2001, Rudolph2003, Rudolph2005, ChenWu2003}. This method is based on the \textit{realignment map}, which permutes the coefficients of the density matrix $\rho=\sum_{ijkl} \rho_{ijkl} |ik \rangle\langle jl| \mapsto \sum_{ijkl} \rho_{\pi(ijkl)} |ik \rangle\langle jl|$, for the index permutation $\pi(ijkl)= (ikjl)$. For all separable states, this permutation results in an operator $\rho_{\pi(ijkl)}$ satisfying $\| \rho_{\pi(ijkl)}\| \leq 1$. The partial transposition is a very similar operation, the only difference being that the permutation $\pi(ijkl)=(ijlk)$ is considered. Generally, the PPT and CCNR criteria are hence used as complementary to each other in the sense that none of them detects all possible entangled states, but some states detected by PPT are not detected by CCNR and vice versa.


\textbf{Beyond linear witnesses}.\ To improve over linear witnesses, a very useful method which allows for implementations in experiments (see, e.g.,~\cite{Zhao-Wolf2019}) makes use of {\it local uncertainty relations} (LURs). The idea to derive entanglement criteria by means of LURs has some analogies with the original EPR-Bell approach in the sense that it considers pairs of non-commuting single party observables, say $(A_1, A_2)$ for party A and $(B_1 , B_2)$ for party B. Since the $A_i$ do not commute with each other, their uncertainties cannot both be zero simultaneously. The same is true for the $B_i$. However, in the joint system, the uncertainties of the \textit{collective observables} $M_i = A_i \otimes \mathbbm{1} + \mathbbm{1} \otimes B_i$ can both vanish at the same time, provided that the state is entangled.

A powerful and at the same time instructive example is given in terms of the variance $(\Delta A)^{2}_{\rho} = \tr(A^{2}\rho)- \tr(A\rho)^{2}$. The sum $(\Delta A_{1})^{2}_{\rho\subtiny{0}{0}{A}} + (\Delta A_{2})^{2}_{\rho\subtiny{0}{0}{A}} \geq U\subtiny{0}{0}{A}$ must have a non-zero lower bound $U\subtiny{0}{0}{A}>0$ for all single-party states $\rho\subtiny{0}{0}{A}$ whenever the two observables do not commute. Similarly, $(\Delta B_1)^2_{\rho\subtiny{0}{0}{B}} + (\Delta B_2)^2_{\rho\subtiny{0}{0}{B}} \geq U\subtiny{0}{0}{B}$ for all $\rho\subtiny{0}{0}{B}$. Thus, by simple concavity arguments one can prove that $(\Delta M_1)^2_{\rho\subtiny{0}{0}{AB}} + (\Delta M_2)^2_{\rho\subtiny{0}{0}{AB}} \geq U\subtiny{0}{0}{A}+ U\subtiny{0}{0}{B}$ must hold for all separable states $\rho\subtiny{0}{0}{AB} = \sum_k p_k (\rho\subtiny{0}{0}{A} \otimes \rho\subtiny{0}{0}{B})_k$~\cite{Hofman2003, GuehneMechlerTothAdam2006, ZhangNhaZhangGuo2010, SchwonnekDammeierWerner2017}. This method hence combines two conceptual features: first, the LURs themselves \textemdash representing a trade-off between information about different complementary (non-commuting) observable quantities \textemdash and second, the fact that those (non-linear) quantities are either concave or convex. Thus, analogous reasoning can be applied to other quantifiers of uncertainty, such as, e.g., the {\it Quantum Fisher Information} (QFI)~\cite{BraunsteinCaves1994}, which has been introduced in the context of quantum metrology and proven to be related to metrological applications of entanglement~\cite{TothApellaniz2014}. Also, LURs in the form of a product of uncertainties (e.g., variances) can be used, although requiring a somewhat more complicated mathematical treatment, to derive entanglement criteria resembling Heisenberg uncertainty relations in their original formulation~\cite{Reid1989, LangeEtAl2018, ReidDrummondBowen2009}.

It is also worth mentioning here that all non-linear entanglement witnesses arising from sums of variances can be cast in a compact form in terms of the covariance matrix $\Gamma_{ij}(\rho)=\tfrac{1}{2} \langle g_i g_j + g_j g_i\rangle_\rho-\langle g_i \rangle_\rho \langle g_j\rangle_\rho$ of a local basis of observables. The resulting {\it covariance matrix criterion} (CMC)~\cite{GuehneHyllusGittsovichEisert2007, GittsovichGuehneHyllusEisert2008, GittsovichHyllusGuehne2010} was proven to be necessary and sufficient for the special case of two qubits, provided that one makes use of {\it local filterings} that map the state to its filtered normal form (FNF) $\rho \mapsto \rho_{\rm FNF}:=(F\subtiny{0}{0}{A} \otimes F\subtiny{0}{0}{B}) \rho (F\subtiny{0}{0}{A} \otimes F\subtiny{0}{0}{B})^\dagger$ such that $\rho_{\rm FNF}=\frac{1}{4}(\mathbbm{1}_{4}+\sum_{i,j=x,y,z}t_{ij}\sigma_i\otimes\sigma_j)$, where $\sigma_k$ are the Pauli matrices. For local dimensions larger than 2, the CMC can in principle be evaluated using semi-definite programs, but in its general form this still proves to be a difficult task, even for bipartite systems.


\textbf{Bounding witnessed entanglement}.\ When using an approach based on witnesses, one is of course also interested in quantitative statements about the detected entanglement based on the data of the (preferably) few measurements required for the witness itself. A simple but general method to compute lower bounds on convex functions of quantum states $E(\rho)$ (such as entanglement measures) using only few expectation values is based on \emph{Legendre transforms}~\cite{GuehneReimpellWerner2007, EisertBrandaoAudenaert2007}. In this context, let us define such a transform as $\hat E(W):= \sup_\rho [\tr(W \rho)-E(\rho)]$, where the supremum is taken over quantum states $\rho$. Note that for a given convex function $E(\rho)$, the quantity $\hat E(W)$ only depend on the chosen witness $W$. Then, a tight lower bound on $E(\rho)$ for the underlying (unknown) system state $\rho$ is obtained via another Legendre transformation, which leads to
\vspace*{-1mm}
\begin{align}
E(\rho) \geq \sup_\lambda [\lambda \tr(W \rho)-\hat E(\lambda W)] ,
\end{align}
where $\lambda$ is a real and $\tr(W \rho)$ is obtained from measurements. The applicability of this technique largely depends on whether $\hat E(W)$ (and hence $E(\rho)$ for given $\rho$) can be efficiently computed, but has turned out to be a powerful tool to quantify multipartite entanglement based on uncertainty relations~\cite{SoerensenMoelmer2001,vitagliano16,VitaglianoPlanar,MartyVitagliano}. Another option is a direct construction of witnesses that themselves have a natural connection between their expectation value and a suitably chosen entanglement measure~\cite{MaChenChenSpenglerGabrielHuber2011, WuKampermannBrussKloecklHuber2012, HuberDeVicente2013}.


\textbf{Measurement strategies}.\ The previous discussion on bipartite entanglement already showcases one of the central challenges for experimental verification: Methods for entanglement quantification and detection are available in abundance, but often defined in a formal way. Some allude to observable quantities, some to maps on density matrices, others to positive operator-valued measures (POVMs). Identifying the most suitable and efficient practical method for a specific experimental setup at hand is hence not straightforward. For instance, the types of measurements that can be most easily (or at all) implemented depend on the experimental platform, and their identification and comparison may be obfuscated by varying terminologies. A consistent challenge across all platforms and paradigms is the exponential number of potential measurements that could be required for the desired task. Moreover, how many measurements are needed is often counted in different terms, such as the number of global settings, the number of local settings, the number of observables, or the number of density matrix elements. To provide a comparative overview over the complexity of different detection methods let us therefore give more precise definitions, briefly review some practical methods of data acquisition, and identify which tests work well with what type of data.

Formally all measurements can be described by POVMs, i.e., sets of positive semi-definite operators $M_i\geq 0$ with the property $\sum_{i=1}^{m} M_{i}=\mathbbm{1}_d$, where $m$ is the number of distinguishable outcomes labelled by `$i$'. A special case is the projective measurement (PM), where $M_i=\ket{v_{i}}\!\!\bra{v_{i}}$ for all $i$ and $m=d$. Each POVM can be thought of as a PM on a larger system, and most experimental implementations indeed work directly with PMs. Repeated PMs allow estimating the expectation values $\tr(\rho M_{i})=\bra{v_{i}}\rho\ket{v_{i}}$, i.e., a complete set of \emph{diagonal} density matrix elements w.r.t. a specific basis $\{v_{i}\}_{i}$, and in turn, the expected values of all observables of the form $O=\sum_{i}\lambda_{i}\ket{v_{i}}\!\!\bra{v_{i}}$. For example, for the simple case of a single qubit, the density matrix can be represented via the \emph{Bloch decomposition} $\rho = \tfrac{1}{2}(\mathbbm{1}+\mathbf{a}\cdot\mathbf{\sigma})$. The diagonal elements of $\rho$ can be obtained via PMs in the computational basis $\{\ket{0},\ket{1}\}$. In the Bloch picture, this corresponds to measuring the $\sigma_z$ operator: $a_z \equiv \tr(\sigma_z\rho)=\bra{0}\rho\ket{0}-\bra{1}\rho\ket{1}$. The off-diagonal elements of $\rho$ cannot be obtained directly from projective measurements in the $\{\ket{0},\ket{1}\}$ basis. Instead, these are indirectly obtained by making projective measurements in the conjugate bases $\{\ket{+},\ket{-}\}$ and $\{\ket{+i},\ket{-i}\}$, where $\ket{\pm} = \tfrac{1}{\sqrt{2}}\bigl(\ket{0}\pm\ket{1}\bigr)$ and $\ket{\pm i} = \tfrac{1}{\sqrt{2}}\bigl(\ket{0}\pm i\ket{1}\bigr)$. Going back to the Bloch picture, this is equivalent to measuring the $\sigma_x$ and $\sigma_y$ operators:  $a_x \equiv \tr(\sigma_x\rho)=\bra{+}\rho\ket{+}-\bra{-}\rho\ket{-}$ and $a_y \equiv \tr(\sigma_y\rho)=\bra{+i}\rho\ket{+i}-\bra{-i}\rho\ket{-i}$. Alternatively, we can look at the example of a specific off-diagonal element $\bra{0}\rho\ket{1}$, whose real and imaginary parts are obtained in the following manner: $\operatorname{Re}\bigl(\bra{0}\rho\ket{1}\bigr) = \tfrac{1}{2}\bigl(\bra{+}\rho\ket{+}-\bra{-}\rho\ket{-}\bigr)$, and $\operatorname{Im}\bigl(\bra{0}\rho\ket{1}\bigr) = \tfrac{1}{2}\bigl(\bra{+i}\rho\ket{+i}-\bra{-i}\rho\ket{-i}\bigr)$.


\textbf{Local vs. global}.\ For the purpose of making statements about entanglement it is useful to distinguish between different types of PMs. Most importantly, one differentiates between \emph{local} and \emph{global} measurement bases (or observables), depending on whether the basis vectors $\ket{v_{i}}$ are product states $\ket{v_{i}}\subtiny{0}{0}{AB} =\ket{u_{i}}\subtiny{0}{0}{A} \otimes \ket{w_{i}}\subtiny{0}{0}{B}$ w.r.t. to the chosen bipartition $A|B$, or not. Here, the choice of basis $\{\ket{v_{i}}\subtiny{0}{0}{AB}\}_{i}$ is referred to as a global setting, whereas bases $\{\ket{u_{i}}\subtiny{0}{0}{A}\}_{i}$ or $\{\ket{w_{j}}\subtiny{0}{0}{B}\}_{j}$ are called local settings. In the standard scenario for quantum communication, whenever the constituents of the quantum system are spatially separated, local (product basis) measurements are the only possible measurements. In this case, detection, certification, or quantification of entanglement requires the measurement of (at least some) off-diagonal density matrix elements. As in the single-qubit example above, these can be obtained by measurements of diagonal matrix elements of specific (product) bases conjugate w.r.t. the original basis. Alternatively, it is often useful to work directly with a local operator basis. That is, the Bloch picture can be extended to $d$-dimensional systems (qudits) any number of parties in terms of a \emph{generalised Bloch decomposition}~\cite{BertlmannKrammer2008} by expanding a quantum state in a basis of suitable matrices $g_{i}$, i.e., $\rho=\sum_{i_{1},i_{2},\ldots,i_{n}=0}^{d^{2}-1}\rho_{i_{1}i_{2}\ldots i_{n}}g_{i_{1}}\otimes g_{i_{2}}\otimes\ldots\otimes g_{i_{n}}$. For instance, for two qudits and an operator basis that includes the identity, one has
\begin{align}
        \rho &=\tfrac{1}{d^{2}}\bigl(\mathbbm{1}_{d^{2}}
        +\vec{v}\subtiny{0}{0}{A}\, \vec{\sigma}\otimes\mathbbm{1}_{d}
        +\vec{v}\subtiny{0}{0}{B}\, \mathbbm{1}_{d}\otimes\vec{\sigma}
        +\sum_{i,j}t_{ij}\sigma_i\otimes\sigma_j\bigr),\nonumber
\end{align}
where $\{\sigma_i\}_{i}$ is a basis of the $SU(d)$ algebra. The Bloch coefficients themselves are obtained as expectation values of local observables, $t_{ij}=\expval{\sigma_i\otimes\sigma_j}_\rho$, making the Bloch basis a convenient expression of quantum states only in terms of results of local measurements instead of abstract density matrix elements. While in general there exist $d^{2}-1$ orthogonal generators of $SU(d)$, requiring a large amount of observables to be measured for tomographic purposes (i.e., the $g_{i}$ generally do not have full rank), most of them can be represented via dichotomic operators and are thus often easier to implement than multi-outcome measurements. In contrast to any local measurements, probes interacting with multiple constituents of the system simultaneously or global observables whose eigenstates do not factorise (e.g., the magnetisation) can give rise to entangling measurements. These measurements are inherently global and the individual detector events can be used directly to estimate the correllators necessary for measuring entanglement witnesses. This is particularly relevant experimentally when the number of involved parties becomes very large, say $n \sim 10^{3}-10^{12}$ or larger, in which case a reconstruction of the full density matrix is prevented by the extremely large number of required measurements. At the same time it is typically possible to measure level populations and consequently infer moments of $N$-particle collective operators like $J_k=\sum_{i=1}^{n} j_k^{(i)}$. Such quantities are in turn directly related to inter-particle correlations, potentially providing information about entanglement.


\textbf{Multi-outcome vs. single-outcome}.\ Measurements in any basis may be classified by the method by which (relative frequencies of) different measurement outcomes are recorded. On the one hand, in \emph{multi-outcome} measurements, the interaction of a measurement device with a single copy of the measured system described by $\rho$ provides one of several (ideally one of $d$) different outcomes `$i$' associated with the projection into $\ket{v_{i}}$. That is, the detector event may fall into one of $d$ categories that can be distinguished by the experimenter. After $N$ such rounds of multi-outcome measurements, each resulting in one detector event, the outcome `$i$' is obtained $S_{i}$ times, such that $\sum_{i=1}^{d}S_{i}=N$, and the expected value of $M_{i}$ is estimated to be $\tr(M_{i}\rho)\approx S_{i}/N$.

In \emph{single-outcome} measurements, on the other hand, filters are used to select only one particular outcome `$i$', for which the detector (e.g., a photo detector placed behind a polarisation filter) responds with a `click'. In principle, one may think of a `no click' event as a second outcome, but this only works if the imminent event is heralded. A much simpler alternative is usually to collect the number $S_{i}$ of `clicks' in the filter setting `$i$' during some fixed integration period and again associate $\bra{v_{i}}\rho\ket{v_{i}}\approx S_{i}/N$ with $N=\sum_{i=1}^{d}S_{i}$ for the chosen orthonormal basis $\{\ket{v_{i}}\}_{i}$. For non-orthonormal bases, this approach can still be used with minor modifications~\cite{BavarescoEtAl2018}. Crucially, the data corresponding to a $d$-outcome measurement can also be obtained from $d$ individual single-outcome measurements. In principle, this also applies to local measurements. For instance, (diagonal) density matrix elements w.r.t. the product basis $\{\ket{u_{i}}\subtiny{0}{0}{A} \otimes \ket{w_{j}}\subtiny{0}{0}{B}\}_{i,j=1}^{d}$ in a $d\times d$-dimensional Hilbert space can be obtained using $d^{2}$ pairs of \emph{local filter} settings, provided that local detection events for filter settings `$i$' and `$j$' fall within a sufficiently close time interval to be combined to `coincidences' $C_{i_{A}j_{B}}$. More generally, for $n$ parties, temporal coincidence allows to associate the localised single events at $n$ detectors into coincidences $C_{i_{1}i_{2}\ldots i_{n}}$ and global density matrix elements
\begin{align*}
    \bra{i_{1}i_{2}\ldots i_{n}}\rho\ket{i_{1}i_{2}\ldots i_{n}} &= \frac{C_{i_{1}i_{2}\ldots i_{n}}}{\sum_{i_{1},i_{2},\ldots,i_{n}} C_{i_{1}i_{2}\ldots i_{n}}}.
\end{align*}


\begin{smallboxtable}{Minimal number of measurement settings}{ms}
\begin{center}
\renewcommand{\arraystretch}{1.4}
\begin{tabular}{|>{\centering\arraybackslash}m{0.5\textwidth}||>{\centering\arraybackslash}m{0.15\textwidth}|>{\centering\arraybackslash}m{0.15\textwidth}|>{\centering\arraybackslash}m{0.15\textwidth}|
}
\hline
\begin{small} \textbf{Quantifier} \end{small} 	& \, {FST} \, &  \,$F(\rho,\Phi)$ \,	& \, $\tr(\rho W)$ \,\\
\hline
\begin{small} {Global Observables} $O$ \end{small}  & \,  $d^n+1$   \, 	& $1$			& $1$ \\
\hline
\begin{small} {Collective observables} $O=\sum_i o_i\otimes \mathbbm{1}_{\overline{i}}$ \end{small}  & \, $\leq (d+1)^n$   \, 	& n.a.			& $2$\,[$1^{*}$]  \\
\hline
\begin{small} {Bi-product bases (local MUB)} $\operatorname{eig}(O)=\{\bigotimes_{i=1}^{2}\ket{|v_{i}}\}$\end{small}  & \,  $(d+1)^2$   \, 	& $d+1$\, [$2^{**}$]			& $2$ \\
\begin{small} {Product bases (local MUB)} $\text{eig}(O)=\{\bigotimes_{i=1}^n|v_i\rangle\}$\end{small}  & \,  $(d+1)^n$   \, 	& $\leq (d+1)^n$			& $2$ \\
\hline
\begin{small} {Bi-Product Bloch bases} $O=\sigma_1^{j_1}\otimes\sigma_2^{j_2}$\end{small}  	& \,  $(d^2-1)^2$	\, & $d^2-1$		& $2$ \\
\hline
\begin{small} {Product Bloch bases} $O=\bigotimes_{i=1}^n\sigma_i^{j_i}$ \end{small}  	& \,  $(d^2-1)^n$	\, & $\leq (d^2-1)^n$		& $2$ \\
\hline
\begin{small} {Local filters} $O=\bigotimes_{i=1}^n\ket{v_i}\!\!\bra{v_i}$\end{small}  	& \,  $(d(d+1))^n$	\, & $(d+1)d^n$		& $2d^n$\, [$2^{***}$] \\
\hline
\end{tabular}\\
\end{center}
\small{
The table shows the minimal number of required measurement settings in terms of different commonly used quantifiers to perform full state tomography (FST), optimal estimation of fidelity w.r.t. pure target states $\Phi$ [$F(\rho,\Phi):=\langle\phi|\rho|\phi\rangle$], or to evaluate an entanglement witness for $2$ or $n$ $d$-dimensional subsystems. Global observables can be used for optimal tomography based on mutually unbiased bases (MUBs)~\cite{WoottersFields1989}, to estimate the fidelity via the observable $O=\Phi=|\phi\rangle\langle\phi|$, or directly represent entanglement witnesses $O=W$. Collective observables are (weighted) averages of single-party observables that can be used to witness entanglement via their second moments [$^{*}$ the second moments of a single observable given by a weighted sum of local observables, i.e., where terms are local but may act nontrivially on more than one subsystem, e.g., for an interaction Hamiltonian, are sufficient to certify entanglement~\cite{DowlingDohertyBartlett2004,tothpra05}]. Local (bipartite or $n$-partite) measurements in MUBs (or tilted bases~\cite{BavarescoEtAl2018}) can be used for local tomography and direct fidelity estimation ($^{**}$ or for certifying a lower bound with only two product bases~\cite{BavarescoEtAl2018}). Determining the coefficients of the Bloch decomposition requires the measurement of all $d^{2}-1$ local Bloch vector elements in every possible combination. Two anti-commuting operators, however, are already sufficient for constructing entanglement witnesses in bipartite~\cite{AsadianErkerHuberKloeckl2016} and multipartite systems~\cite{TothGuehne2005a,LaskowskiMarkiewiczPaterekZukowski2011}. Post-selecting coincidence counts in a single-outcome scenario (i.e., filtering) requires every possible projection on a tomographically complete set of states, i.e., $d^n$ measurement settings per single basis ($^{***}$ although it is possible to detect entanglement without knowing any density matrix element~\cite{Tiranov-Gisin2016}.
}
\end{smallboxtable}


\textbf{Statistical error and finite data}.\ The discussion above illustrates that the number of measurement settings required for entanglement tests does not just depend on the chosen theoretical method, but also on what is counted: local or global bases/operators; filter settings (single-outcome), dichotomic observables (two outcomes, e.g., for Bloch decompositions), or multi-outcome measurements. However, regardless of the method used, each single measurement setting still requires a number of repetitions of individual measurements to ensure the desired statistical confidence in the result. That is, the association $\tr(M_{i}\rho)\approx S_{i}/N$ is exact only in the limit of infinitely many repetitions and any real experiment using finitely many measurements may only estimate probabilities or expected values from frequencies of occurrence of certain measurement outcomes. The confidence in these estimates is then guaranteed by a sufficiently large sample size (number of repetitions) by way of the central limit theorem and Hoeffding's inequality. How many samples can be taken with reasonable effort and time largely depends on the specific experimental setup. For instance, while many thousands of coincidences can be recorded every second in photonic setups used in communications and the resulting statistical error can both easily be computed and does not heavily influence the conclusions drawn, state preparation in other systems is often tedious and not straightforwardly repeatable. In such scenarios, statistical errors and sufficiently narrow confidence intervals become prominent challenges that have to be addressed. Certifying entanglement with finite data was first addressed by~\cite{BlumeKohoutYinVanEnk2010} with simulated two-qubit data, but similar reasoning also applies to methods directly aimed at state estimation~\cite{FlammiaLiu2011, AolitaGogolinKlieschEisert2015}. In this context Ref.~\cite{SchwemmerKnipsRichartMoroderKleinmannGuehneWeinfurter2015} also provides a cautionary tale against density matrix reconstruction techniques, as the negligence of errors can lead to systematic overestimation of entanglement and underestimation of fidelity (maximum likelihood reconstructions have thus recently been deemed inadmissible for fidelity estimation~\cite{FerrieBlumeKohout2018}). In general, different measurement techniques come at different experimental cost for entanglement estimation or state tomography. This cost can be quantified in the number of states needed for achieving statistical certainty (see, e.g., Ref.~\cite{PallisterLindenMontanaro2018} for optimal strategies in the bipartite case). Nonetheless, in case that sufficiently many repetitions for meaningful statistics are possible (e.g., for down-converted photons) the number of different measurement bases/settings remains the principal measure of efficiency. An overview over this figure of merit for the most common measurement strategies is shown in Tab.~\ref{table:ms}, while some examples for entanglement detection methods are shown in Tab.~\ref{table:examples}.


\begin{smallboxtable}{Examples for entanglement detection methods}{examples}
\begin{center}
{\renewcommand{\arraystretch}{1.5}
\begin{tabular}{|>{\centering\arraybackslash}m{0.08\textwidth}||>{\centering\arraybackslash}m{0.3\textwidth}|>{\centering\arraybackslash}m{0.32\textwidth}|>{\centering\arraybackslash}m{0.25\textwidth}|
}
\hline
\begin{small} \textbf{Method} \end{small} 	
    & Witness  $\text{Tr}(\rho W)\geq0$
    & Nonlinear Witness $f(\rho)\geq0$
    & Positive Map $\Lambda[\rho]\geq 0$\\
\hline
\begin{small} {Two qubits}  \end{small}
    & $-\,\operatorname{Re}\bigl(\bra{00}\rho\ket{11}\bigr)$
        $+\,\frac{1}{2}\bigl(\bra{01}\rho\ket{01}$  $+\,\bra{10}\rho\ket{10}\bigr)$ 	
    & $\sqrt{\bra{01}\rho\ket{01}\bra{10}\rho\ket{10}}-|\bra{00}\rho\ket{11}|$
    & $\rho\mapsto\rho^{T_A}$ \\
\hline
\begin{small} {Two qutrits}  \end{small}
    & $\frac{2}{3}-\tfrac{2}{3}\operatorname{Re}\bigl(\bra{00}\rho\ket{11}$
        $+\bra{00}\rho\ket{22}+\bra{11}\rho\ket{22}\bigr)$
        $-\tfrac{1}{3}\bigl(\bra{00}\rho\ket{00}+\bra{11}\rho\ket{11}$
        $+\bra{22}\rho\ket{22}\bigr)$	
    & $\det\bigl(M\bigr)$\ with
        $M_{ij}=\tfrac{1}{2}\Bigl[2\delta_{ij}\bra{i}
        \rho\subtiny{0}{0}{B}\ket{j}-\bra{ii}\rho\ket{jj}\Bigr]$
    & $\rho\mapsto$
        $\mathbbm{1}\subtiny{0}{0}{3}\otimes\rho\subtiny{0}{0}{B}-\tfrac{1}{2}\rho\subtiny{0}{0}{AB}$ \\
\hline
\begin{small} {Three qubits}  \end{small}
    & $\operatorname{Re}\bigl(\bra{000}\rho\ket{111}\bigr)$
        $-\tfrac{1}{2}\bigl(\bra{001}\rho\ket{001}+\bra{110}\rho\ket{110}\bigr)$
        $-\tfrac{1}{2}\bigl(\bra{010}\rho\ket{010}+\bra{101}\rho\ket{101}\bigr)$
        $-\tfrac{1}{2}\bigl(\bra{100}\rho\ket{100}+\bra{011}\rho\ket{011}\bigr)$
    & $-|\bra{000}\rho\ket{111}|$
        $+\sqrt{\bra{001}\rho\ket{001}\bra{110}\rho\ket{110}}$
        $+\sqrt{\bra{010}\rho\ket{010}\bra{101}\rho\ket{101}}$
        $+\sqrt{\bra{100}\rho\ket{100}\bra{011}\rho\ket{011}}$
    & $\rho\mapsto\mathbbm{1}+\sigma_x^A\rho^{T_A}\sigma_x^A+$
    $\sigma_x^B\rho^{T_B}\sigma_x^B+\sigma_x^C\rho^{T_C}\sigma_x^C$\\
\hline
\end{tabular}\\
}
\end{center}
\begin{small}
The table presents some illustrative examples for linear and nonlinear (in $\rho$) witnesses (negative values detect), positive (but not completely positive) maps (resulting non-positive operators detect) detecting bipartite entanglement for two-qubits, maximal entanglement dimensionality (i.e., Schmidt number $3$) for $2$ qutrits, and genuine multipartite entanglement for $3$ qubits. All of these examplary techniques detect entanglement/Schmidt number/GME for the generalised state $|\psi\rangle=\tfrac{1}{\sqrt{d}}\sum_{i=0}^{d-1}\ket{i}^{\otimes n}$ for $(n,d)=(2,2)$, $(2,3)$, and $(3,2)$, respectively.
\end{small}
\end{smallboxtable}


\section*{Contemporary challenges: High-dimensional entanglement}

\textbf{Entanglement dimensionality}.\ High-dimensional Hilbert spaces enable an encoding of more bits per photon than in the two-dimensional case and thus promise increased communication capacities over quantum channels~\cite{MirhosseiniEtAl2015} and an increased robustness to noise~\cite{Ecker-Huber2019}. However, if the security of these channels is to be ensured by entanglement, a major challenge presents itself in the certification of high-dimensional entanglement. There, the principal goal is to certify entanglement with as few measurements as possible, without introducing unwarranted assumptions that may lead to exploitable loopholes in the certification. In this context matrix completion techniques~\cite{Tiranov-Gisin2017, Martin-Gisin2017}, semi-definite programs~\cite{Martin-Gisin2017, Steinlechner-Ursin2017}, uncertainty relations~\cite{SchneelochHowland2018} and mutually unbiased bases~\cite{ErkerKrennHuber2017, TascaSanchezPieroWalbornRudnicki2018, BavarescoEtAl2018} have proven to provide versatile tools for quantifying high-dimensional entanglement in different contexts.

The canonical witnesses for known target states $\ket{\psi\subtiny{0}{0}{\mathrm{T}}}$ shown in Box~\ref{boxbox:witnesses} can readily be generalised to detect high-dimensional entanglement in the same fashion. One defines $W_k:=\sum_{i=1}^k\lambda_{i}^2\mathbbm{1}-\ket{\psi\subtiny{0}{0}{\mathrm{T}}}\!\!\bra{\psi\subtiny{0}{0}{\mathrm{T}}}$, where $\sum_{i=1}^k\lambda_{i}^2$ denotes the sum over the $k$ largest squared Schmidt coefficients of the target state~\cite{PianiMora2007}. While this witness faithfully certifies high-dimensional entanglement of any pure target state, it is decomposable (i.e., detects only NPT states) and features a weak resistance to noise. On the other hand, it only requires an estimate of the target state fidelity which can be efficiently obtained with few measurements~\cite{PallisterLindenMontanaro2018, BavarescoEtAl2018}.

High-dimensional entanglement can also be ascertained using suitable quantitative measures. For instance, certifying an entanglement of formation beyond $\log_2(k)$, also implies $(k+1)$-dimensional entanglement. Alternatively, high-dimensional entanglement may as well be quantified directly by the g-concurrence~\cite{Gour2005}, bounds for which can be obtained from non-linear witness operators~\cite{SentisEltschkaGuehneHuberSiewert2016}.

From a local Hilbert space perspective, multiple copies of entangled qubit pairs can be considered as equivalent to high-dimensionally entangled systems. However, this equivalence breaks down for distributed quantum systems. Genuine high-dimensionally entangled systems can feature correlations in principle unattainable by multiple copies of two-qubit entangled states~\cite{KraftRitzBrunnerHuberGuehne2018}, which has recently also been used in a photonic experiment to verify genuine high-dimensional entanglement~\cite{GuoHuLiuHuangLiGuo2018}.

Besides practical challenges, many questions still remain concerning the mathematical structure of high-dimensional entanglement. While PPT entanglement is known to generically occur in high-dimensional Hilbert spaces~\cite{SzarekWernerZyczkowski2011}, few techniques are known for constructing corresponding witnesses (or, dual to that problem, non-decomposable $k$-positive maps~\cite{HorodeckiMPR1996}). Even among PPT states, high-dimensional entanglement is generic~\cite{HuberLamiLancienMuellerHermes2018}, but at the same time not maximal~\cite{SanperaBrussLewenstein2001}.\\


\begin{figure}[ht!]
    \begin{center}
    \includegraphics[width=0.5\textwidth]{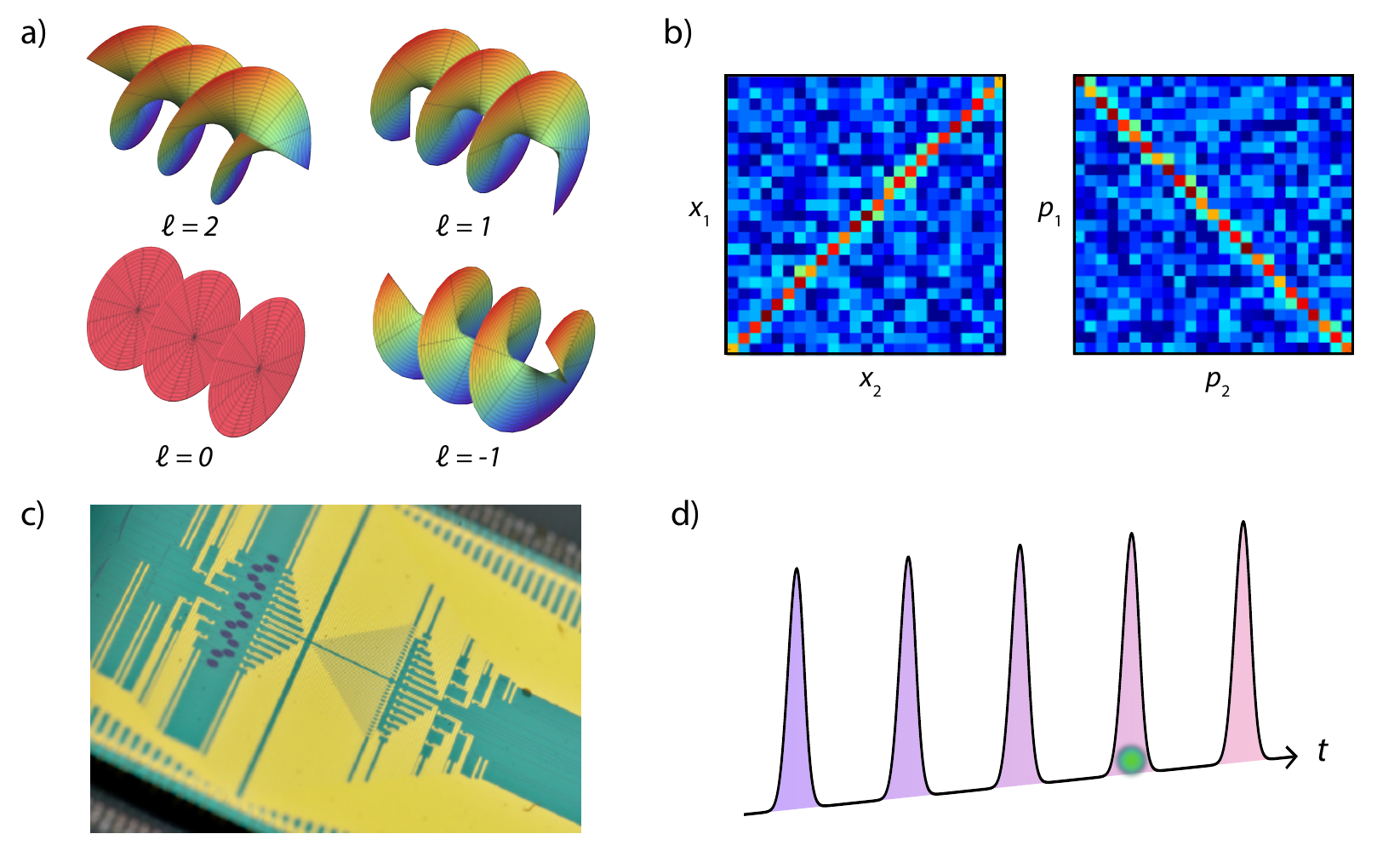}
    \end{center}
    \caption{High-dimensional entanglement has been realised in several photonic degrees-of-freedom. a) Orbital angular momentum (OAM) is one such discrete quantum property that is carried by light beams with a helical phase structure, where the OAM quantum number $\ell$ denotes the direction and number of ``twists" within one wavelength. b) Photons entangled in transverse position-momentum (x-p) exhibit strong correlations and anti-correlations that can be observed on sensitive single-photon cameras, as seen here in simulated data. c) Photonic integrated circuits offer yet another platform for the realisation of high-dimensional quantum states, where information is encoded in different paths of a circuit. The image shows a $16\times16$-dimensional integrated circuit used recently in a test of generalised Bell-type inequalities~\cite{WangEtAl18} (image courtesy of Jian Wei Wang, copyright American Association for the Advancement of Science (AAAS)). d) High-dimensional quantum states can also be encoded in the arrival time of a photon, where a photon can occupy one of several time-bins.}
    \label{fig:PhotonHD}
\end{figure}

\textbf{Platforms for high-dimensional entanglement}.\ Photonic systems, which are inherently multi-mode in the temporal and spatial degrees-of-freedom, naturally lend themselves to the creation and measurement of high-dimensional entanglement, see Fig.~\ref{fig:PhotonHD}. In addition to photons, several other physical platforms exist that offer promising paths towards the controlled realisation and investigation of high-dimensional entanglement. However, the choice of which entanglement quantifier to use from the many discussed above often comes down to what kind of tools are available to the experimenter for making measurements on their quantum system of choice. Here, we discuss some landmark experiments and accompanying theoretical techniques used for demonstrating high-dimensional entanglement of two photons in their orbital angular momentum (OAM), transverse spatial position-momentum, time-frequency, and path degrees-of-freedom (DOFs). We further outline promising developments in the creation and measurement of high-dimensional quantum states of Caesium atoms, superconducting qubits, and Nitrogen-vacancy centers, all of which could serve as interesting future platforms for creating high-dimensional entanglement.

High-dimensional entanglement of two photons in the spatial or temporal degrees of freedom usually results from the conservation of energy and momentum in a second-order non-linear process such as spontaneous parameteric down-conversion (SPDC). This process entails the annihilation of one pump photon with energy $\hbar\omega$ and zero OAM in a non-linear crystal, resulting in the creation of two daughter photons with energy $\frac{1}{2}\hbar\omega$. While formally the dimension of the Hilbert space relating to modal properties is infinite, only finitely many modes will be populated significantly. Thus the effective dimensionality of the resulting two-photon state depends on the spectral and spatial properties of the pump beam, as well as on the phase-matching function governing the non-linear process. For example, the pump beam width and the length of the non-linear crystal determine the dimensionality of an OAM-entangled state~\cite{MiattoGiovanniniRomeroFrankeArnoldBarnettPadgett2012}.

Some of the first demonstrations of high-dimensional entanglement were performed with photons entangled in their OAM, which is a discrete quantum property resulting from a spatially varying amplitude and phase distribution~\cite{AllenBeijersbergenSpreeuwWoerdman1992, KrennMalikErhardZeilinger2017}. This type of entanglement was first demonstrated with Schmidt number $d_{\rm{ent}}=3$ in an experiment that measured a generalised Bell-type (Collins-Gisin-Linden-Massar-Popescu~\cite{CollinsGisinLindenMassarPopescu2002}) inequality with single-outcome, holographic projective filters that allowed the measurement of coherent superpositions of OAM at the single photon level~\cite{VaziriWeihsZeilinger2002}. In recent years, the development of computer-programmable wavefront-shaping devices such as spatial light modulators (SLMs) have allowed the measurement of OAM-entangled states with ever-increasing dimension~\cite{QassimMiattoTorresPadgettKarimiBoyd2014, BouchardHerreraValenciaBrandtFicklerHuberMalik2018}. Examples of such experiments include the certification of $d_{\rm{ent}}=100$ spatial-mode entanglement with a visibility-based entanglement witness~\cite{KrennHuberFicklerLapkiewiczRamelowZeilinger2014} and $d_{\rm{ent}}=11$ OAM-entanglement with a generalised Bell-type test~\cite{DadaLeachBullerPadgettAndersson2011}, both with certain assumptions on the state. More recently, an assumption-free entanglement witness was implemented with SLMs certifying $d_{\rm{ent}}=9$ OAM-entanglement with only two measurement settings~\cite{BavarescoEtAl2018}.

A natural second basis for observing high-dimensional entanglement is found in the transverse photonic position-momentum degrees-of-freedom. A discretised version of transverse position can be thought of as a ``pixel" basis, which is particularly relevant today with the development of sensitive single-photon cameras. Pixel entanglement was first observed with arrays of three and six fibers~\cite{OSullivanAliKhanBoydHowell2005}, and entanglement was certified by violating the EPR-Reid criterion~\cite{Reid1989} lower-bounding the product of conditional variances in position and momentum: $\Delta^2 (\rho_1-\rho_2)\Delta^2(p_1+p_2)\geq \frac{\hbar^2}{4}$. More recently, electron-multiplying cameras that exhibit a high single-photon detection efficiency have been used to violate the EPR-Reid criterion by very high values, albeit by subtracting a large, uncorrelated background~\cite{EdgarTascaIzdebskiWarburtonLeachAgnewBullerBoydPadgett2012, MoreauDevauxLantz2014}. Other approaches that aim to reduce the number of measurements required to certify position-momentum entanglement have been developed, such as using compressed-sensing techniques to measure such states in a sparse basis~\cite{HowlandKnarrSchneelochLumHowell2016} or employing periodic masks in order to increase photon-counting rates~\cite{TascaRudnickiAspdenPadgettSoutoRibeiroWalborn2018}.

It is important to point out here that in several experimental works, the term ``Schmidt number" is used to define a different concept than the canonical one mentioned in the introduction. This surrogate quantity refers to the inverse purity, which for pure states is related to the Schmidt coefficients via $\operatorname{PR}(\ket{\psi})=\bigl(\sum_i\lambda_i^4\bigr)^{-1}$ and is supposed to roughly quantify the number of local dimensions that relevantly contribute to the observed coincidences. This approach was introduced~\cite{LawEberly2004} to describe pure continuous-variable systems, where the Schmidt rank of pure two-mode squeezed states is infinite while any proper entanglement entropy is still finite (in particular, the inverse purity is the exponential of the R{\'e}nyi-$2$ entropy of entanglement).

The development of silicon integrated photonic circuits presents another versatile platform for high-dimensional entanglement, where quantum states are simply encoded in different optical paths of a circuit. While such circuits have been used extensively for quantum information processing with qubits~\cite{MatthewsPolitiStefanovObrien2009, SansoniSciarrinoValloneMataloniCrespiRamponiOsellame2010}, their first implementation for qutrit entanglement was demonstrated only recently, with integrated multiport devices enabling the realisation of any desired local unitary transformation in a two-qutrit space~\cite{SchaeffPolsterHuberRamelowZeilinger2015}. A more recent experiment certified up to $d_{\rm{ent}}=14$ through the use of nonlinear device-independent dimension witnesses in a large-scale 16-mode photonic integrated circuit, and demonstrated violations of a generalised Bell-type inequality~\cite{CollinsGisinLindenMassarPopescu2002} and the recently developed Salavrakos--Augusiak--Tura--Wittek--Ac{\'i}n--Pironio (SATWAP) inequality~\cite{SalavrakosAugusiakTuraWittekAcinPironio2017} in up to $d_{\rm{ent}}=8$~\cite{WangEtAl18}.

Alongside position-momentum encoding, the time-frequency domain presents yet another powerful platform available for the investigation of high-dimensional entanglement. Early experiments in this direction demonstrated high-dimensional entanglement in photonic time-bins generated by SPDC with a mode-locked, pulsed pump laser~\cite{DeRiedmattenMarcikicZbindenGisin2002, ThewAcinZbindenGisin2004}. A central challenge in certifying time-bin entanglement is measuring coherent superpositions of multiple time-bins. Usually performed with unbalanced interferometers, this method can only measure a single two-dimensional subspace at a time and faces problems of scalability and stability. A recent experiment overcame this problem through the use of matrix completion methods that required only coherent superpositions of adjacent time-bins in order to certify $d_{\rm{ent}}=18$ entanglement with 4.1 ebits of entanglement of formation~\cite{Martin-Gisin2017}. In parallel, experiments certifying high-dimensional frequency-mode entanglement have also been demonstrated, for example by the manipulation of broadband SPDC via SLMs~\cite{BessireBernhardFeurerStefanov2014}, or through electro-optic phase modulation of photons generated via spontaneous four-wave mixing (SFWM) in integrated micro-ring resonators~\cite{KuesEtAl2017, Kues-Morandotti2019}. Finally, multiple photonic degrees-of-freedom can be combined to produce what is referred to as hyperentanglement. This was first demonstrated with photonic OAM, time-frequency, and polarisation, where entanglement was certified in each degree-of-freedom via a Bell CHSH test~\cite{BarreiroLangfordPetersKwiat2005}. More recently, a hyperentangled state of polarisation and energy-time was transmitted over 1.2km of free-space, and high-dimensional entanglement in $d_{\rm{ent}}=4$ was certified via an entanglement witness relating visibility to state fidelity~\cite{Steinlechner-Ursin2017}.

In addition to photonic systems, high-dimensional quantum states have been realised in several matter-based systems, demonstrating their potential as a high-dimensional entanglement platform. For example, the electron and nuclear spins of individual Caesium atoms were recently used for implementing $16$-dimensional unitary transformations via radio frequency and microwave magnetic fields~\cite{AndersonEtAl2015} , resulting in fidelities greater than $0.98$. Control over the first three energy levels of a transmon superconducting circuit was demonstrated with microwave fields via the process of stimulated Raman adiabatic passage (STIRAP)~\cite{KumarVepsaelaeinenDanilinParaoanu2016}. Population was transferred from the ground to the second excited state, without populating the intermediate first excited state. Around the same time, three levels of a Nitrogen vacancy (NV) centre electron spin were used to simulate the bond disassociation energy of a Helium Hydride cation, while the associated nuclear spin was used as a probe qubit for energy readout~\cite{WangEtAl15}. These examples highlight the potential that matter-based systems offer for the field of high-dimensional entanglement, with several new implementations surely on the horizon.

Finally, it is worth mentioning that recent progress has been made on entangling two micromechanical oscillators consisting of nano-structured silicon beams. An optical field was used to excite a single phonon in either of the two oscillators, followed by quantum erasure and post-selection for heralding the entangled state~\cite{RiedingerEtAl2018}. Quantum opto-mechanical systems such as these may provide yet another playground for exploring the types of complex entanglement achieved thus far only with photonic systems.


\section*{Contemporary challenges: Multipartite entanglement}

While multipartite entangled states are ubiquitous in Nature, the controlled generation and manipulation of multipartite entangled states is a principal challenge in current experiments. The appearance of multipartite entangled states across different disciplines comes at no surprise and our review cannot do the complexity of this topic justice. To name just a few, multipartite entanglement forms the basis for quantum networking proposals in quantum communication~\cite{EppingKampermannMacchiavelloBruss2017, PivoluskaHuberMalik2018, RibeiroMurtaWehner2018, BaeumlAzuma2017}, in quantum metrology it is a key resource for beating the standard quantum limit (SQL)~\cite{TothEntanglementMetrologyFisher12}, it is important in quantum error correcting codes~\cite{Scott2004}, appears as a generic ingredient in quantum algorithms~\cite{BrussMacchiavello2011} and appears as the principal resource in measurement-based quantum computation~\cite{RaussendorfBriegel2001, BriegelRaussendorf2001}. The latter two topics motivate the introduction of quantum states representable by graphs~\cite{HeinEisertBriegel2004} or hypergraphs~\cite{RossiHuberBrussMacchiavello2013}. As these are locally equivalent to so-called stabilizer states, the two concepts are often used synonymously and a lot of effort has been invested in certifying entanglement for stabilizer states~\cite{TothGuehne2005b, AudenaertPlenio2005, SmithLeung2006}.

Furthermore, apart from practical, technologically-oriented applications, (multipartite) entanglement has been found closely connected with important physical phenomena, ranging from the physics of many-body systems, to quantum thermodynamics and even quantum gravity. More specifically, in thermodynamics, the typical entanglement of many body systems is a crucial ingredient in reaching thermodynamic equilibrium~\cite{GogolinFriesdorfEisert2015, GogolinEisert2016}, while the growth of entanglement entropies with subsystem areas or volumes is of high importance in the field of condensed matter physics~\cite{AmicoFazioOsterlohVedral2008, EisertCramerPlenio2010, Laflorencie2016}, as well as for approaches to quantum gravity using correspondences between Anti-DeSitter spacetimes with conformal field theories~\cite{Maldacena1999}.

In particular, as nowadays observed in several works, entanglement of thermal states is drastically influenced by quantum criticality: A high degree of entanglement appears in ground states across a quantum phase transition, with a scaling law that depends on the universality class of the transition. So far, theoretical studies in this framework have been devoted to entanglement across bipartitions, especially in ground states, (e.g., quantified by the concurrence of two sites crossing the partition or the von~Neumann entropy of a block), and also to multipartite entanglement in thermal states (e.g., with criteria arising from collective quantities)~\cite{Ma2011Quantum, GuehneToth2009}. In the former case the celebrated \textit{area law} of entanglement for non-critical systems emerged as a major result and the corresponding classification of entangled states as \textit{tensor networks}, a notion closely connected to classical simulability of many-body states~\cite{AmicoFazioOsterlohVedral2008, VerstraeteMurgCirac2008, EisertCramerPlenio2010, Schollwoeck2011, Orus2014, RanTirritoPengChenSuLewenstein2017}.

Thus, given the vastly different types of entanglement that can exist between multiple constituents, it is not surprising that different physical platforms exist for each application, which warrant different approaches to entanglement certification. In the following we hence give an overview of a selection of contemporary theoretical techniques, and highlight their application in exemplary platforms. First, we consider few body systems such as photons and ion traps (see Boxes~\ref{boxbox:GMEphotons} and~\ref{boxbox:GMEions}, respectively), before we move on to another (this time many-body) target platform for current quantum technologies: atomic gases (see Box~\ref{boxbox:coldatombox2}), noting that other promising realisations of multipartite entanglement exist (e.g., using superconducting qubits~\cite{Kelly-Martinis2015, SongEtAl2017, Gong-Pan2019}) but their detailed description goes beyond the scope of this review.


\textbf{Genuine multipartite entanglement}.\ While the previously discussed definitions for entanglement across bipartitions of the systems straightforwardly carry over to the many-particle case, there is a much deeper structure underlying the potential ways in which multipartite systems can be entangled. To unravel this structure, it is instructive to revisit the definition of separability. It is quite intuitive that there exist states of multipartite systems that can be factored into tensor products of multiple parts. This leads to the definition of \emph{k-separable} pure states as $\ket{\Psi_{k-\mathrm{sep}}}:=\bigotimes_{i=1}^k\ket{\Phi_{\alpha_i}}$, where the $\alpha_i\subseteq\{1,2,\cdots,N\}$ refer to specific subsets of systems in the collection of $N$ parties, i.e., $\bigcup_{i=1}^k\alpha_i=\{1,2,\cdots,N\}$ and $\alpha_i\cap\alpha_j=\emptyset \, \forall i\neq j$. States for which $k=N$ are called \emph{fully separable}, as there is no entanglement in the system whatsoever. In the other extreme, states are called \emph{multipartite entangled} if $k=1$, i.e., for all possible partitions of the system one finds entanglement. Considering general (mixed) quantum states adds another layer of complexity to this notion, as k-separability has to be defined as $\rho_{k-\mathrm{sep}}:=\sum_i p_i\ket{\Psi_{k-\mathrm{sep}}^i}\!\!\bra{\Psi_{k-\mathrm{sep}}^i}$,
where each of the $\ket{\Psi_{k-\mathrm{sep}}^i}$ can be separable with respect to a different k-partition. While states with $k=N$ are still fully separable and can be prepared purely by LOCC, the case of $k=1$ is referred to as \emph{genuine multipartite entanglement} (GME). Here, the word ``genuine" emphasises the fact that the state indeed cannot be prepared via LOCC without the use of multipartite entangled pure states. In contrast to the pure state case there hence exist density matrices which are entangled across every partition, and yet do not require multipartite entanglement for their creation.


\begin{WideBoxes}{\rm Genuine multipartite entanglement of photons}{GMEphotons}
The entanglement of more than two photons poses a unique experimental challenge \textemdash\ photons do not interact with each other easily, and higher-order non-linear processes are very inefficient, rendering the direct generation of multi-photon entangled states impractical. Recent experiments have been performed that directly generated three-photon entanglement via cascaded down-conversion, albeit at very low count rates \cite{ShalmHamelYanSimonReschJennewein2013}. Multi-photon entanglement experiments have conventionally relied on an elegant idea introduced by {\.Z}ukowski \textit{et al.}~\cite{ZukowskiZeilingerWeinfurter1995}, where two independent pairs of entangled photons are combined in such a manner as to erase their ``which-source" information. This is illustrated in panel~a) below, where one photon each from two pairs of polarisation-entangled photons are combined at a polarising beam splitter. A polarisation-entangled GHZ state of four photons is obtained by post-selecting on detection events at all four detectors. The first experiment based on the above ideas entangled three photons in their polarisation, showing the presence of a three-photon coherent superposition~\cite{BouwmeesterPanDaniellWeinfurterZeilinger1999}. The same setup was later used to violate a three-particle Mermin inequality, certifying the presence of GME~\cite{PanBouwmeesterDaniellWeinfurterZeilinger2000}.

Subsequent experiments haves since extended this idea of ``entanglement through information erasure", entangling four~\cite{PanDaniellGasparoniWeihsZeilinger2001}, six~\cite{Lu-Pan2007}, eight~\cite{Yao-Pan2012}, ten~\cite{WangEtAl2016}, and most recently,
a record twelve photons~\cite{Zhong-Pan2018} in their polarisation. Due to their low count rates, such experiments have primarily used fidelity-based entanglement witnesses to certify GME. Parallel efforts have aimed at increasing these low probabilistic count rates achieved in multi-photon experiments by tailoring sources to reduce the degree of distinguishability of independent photons~\cite{GraffittiBarrowProiettiKundysFedrizzi2018}. The first experiment extending multipartite entanglement (in any platform) into the high-dimensional regime was recently performed with photonic OAM~\cite{MalikErhardHuberKrennFicklerZeilinger2016}, and applied the ideas of information-erasure to the spatial degree-of-freedom via a specially designed OAM-parity beam splitter~\cite{LeachPadgettBarnettFrankeArnoldCourtial2002}. This experiment hinted at the rich structure that high-dimensional multipartite entanglement can take, by creating a state entangled in $3\times3\times2$ local dimensions (Schmidt rank vector $(3,3,2)^{T}$). Even more recently, the first three-dimensional GHZ state was created with the OAM of photons~\cite{ErhardMalikKrennZeilinger2017}, using the experimental setup pictured in panel~b) below. Interestingly, this setup was found through the use of a computational algorithm \cite{KrennMalikFicklerLapkiewiczZeilinger2016, Melnikov-Briegel2018}, and employed several counter-intuitive techniques departing from the symmetry of the conventional two-dimensional techniques described above. In order to certify high-dimensional GME for these entangled states, a fidelity-based entanglement witness was used for proving that they cannot be decomposed into states of a smaller dimensionality structure~\cite{MalikErhardHuberKrennFicklerZeilinger2016, LeachPadgettBarnettFrankeArnoldCourtial2002}.

%
\begin{center}
    \includegraphics[width=0.8\textwidth,trim={0cm 0.2cm 0cm 0.2cm},clip]{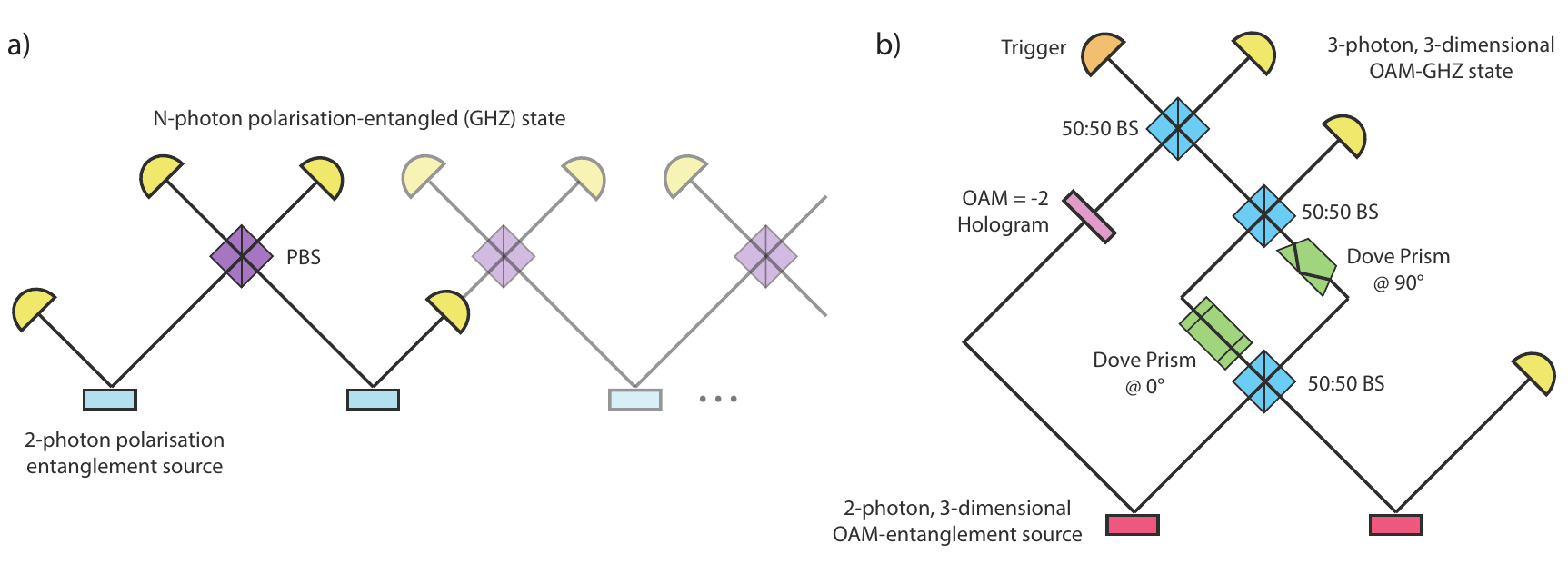}
\end{center}
\begin{small}
The illustration in (a) shows the concept behind entangling more than two photons via information erasure~\cite{ZukowskiZeilingerWeinfurter1995}. Photons from independent pairs of polarisation-entangled photons are combined at a polarising beam splitter (PBS) in such a manner as to erase their ``which-source" information, resulting in a four-photon GHZ state~\cite{PanDaniellGasparoniWeihsZeilinger2001}. The same idea can be generalised to create N-photon entanglement. An experimental setup that extends this idea into the high-dimensional regime~\cite{ErhardMalikKrennZeilinger2017} is shown in (b). Devices such as Dove prisms and spiral phase (OAM) holograms manipulate pairs of photons high-dimensionally entangled in their OAM in order to create a three-particle, three-dimensional GHZ state.
\end{small}
\end{WideBoxes}


\textbf{Entanglement depth}.\ While the above definition reveals one aspect of entanglement in multipartite systems, it is far from a complete characterisation. Take for example the two states $\ket{\psi_1}\otimes\ket{\psi_{234}}$ and $\ket{\psi_{12}}\otimes\ket{\psi_{34}}$. Both are $2$-separable, yet one describes a tripartite entangled system decoupled from a fourth party, while the other represents a pair of independent bipartite entangled states. The concept of \emph{entanglement depth} attempts to capture this distinction, quantifying the number of entangled subsystems in a multipartite state. In the above example the entanglement depths would be three and two respectively. Analogous to GME, the generalisation to mixed states makes use of a contrapositive: A state is called $k$-producible if it can be decomposed as a (mixture of) product of $k$-particle states, i.e., $\rho_{k-\mathrm{prod}}=\sum_i p_i (\rho_{\beta_1} \otimes \dots \otimes \rho_{\beta_M})_i$, where the $\rho_{\beta_m}$ are states of {\it at most $k$ parties}. On the contrary, a state that is not $k$-producible has a depth of entanglement of at least $k+1$~\cite{SoerensenMoelmer2001, GuhneTothBriegel2005}. The two notions of $k$-separability and $k$-producibility are hence quite different, but match in the extremal cases: A fully separable state is also $1$-producible, while a genuine $N$-partite entangled state also has an entanglement depth of $N$ (i.e., it is $N$-producible, but not $(N-1)$-producible). The concept of entanglement depth is particularly useful for systems with very many particles, approaching the thermodynamic limit, since the resulting hierarchy is (somewhat) independent from the total number of particles $N$. Entanglement depth is therefore often used in experiments with atomic ensembles~\cite{2016arXiv160901609P}.


\textbf{Tensor rank \& Schmidt rank vectors}.\ In contrast to the bipartite case, there is no such thing as a Schmidt decomposition (at least not in the same sense) for multipartite systems. That is, not every multipartite state can be written as $\ket{\Psi_N}=\sum_i\lambda_i\ket{i}^{\otimes N}$. Nonetheless, there are two prominent ways to generalise the Schmidt rank for multipartite pure states. One of them is the \emph{tensor rank} $r_{\mathrm{T}}$, which is defined as the minimum number of coefficients $\lambda_i$, such that the state can be written as $\ket{\Psi_N}=\sum_{i=1}^{r_{\mathrm{T}}}\lambda_i\bigotimes_{x=1}^N\ket{v_{i}^{x}}$, such that $\bigotimes_{x=1}^N\scpr{v_i^x}{v_j^x}=\delta_{ij}$. Similar to the Schmidt rank,  $r_{\mathrm{T}}=1$ implies full separability of the state. It is at least NP-hard to determine the tensor rank even for pure states~\cite{Hastad1990}. Moreover, the tensor rank is not additive under tensor products~\cite{ChristandlJensenZuiddam2018} and is known only for very few exemplary multipartite states with particular symmetries~\cite{ChenChitambarDuanJiWinter2010}. One can, however, bound the tensor rank from below by considering the Schmidt ranks with respect to all possible partitions $\alpha_i|\overline{\alpha}_i$, which we denote by $r_{\alpha_i}$ since it is also the rank of the corresponding reduced density matrix $r_{\alpha_i}=\text{rank}(\tr_{\overline{\alpha}_i}\ket{\Psi_N}\!\!\bra{\Psi_N})$. Using this definition it is easy to see that $r_{\mathrm{T}}\geq\max_i r_{\alpha_i}$.
The second generalisation used as an alternative to the tensor rank is the collection of the marginal ranks in the \emph{Schmidt rank vector}~\cite{HuberDeVicente2013} $[\vec{r}\subtiny{0}{0}{\rm{S}}]_i:=r_{\alpha_i}$. Since there are $2^{N-1}-1$ possible bipartitions of the system, this vector has exponentially many components and a state is fully separable if and only if $||\vec{r}\subtiny{0}{0}{\rm{S}}||^2=2^{N-1}-1$, i.e., if every marginal rank is equal to one. While this vector admits different ranks across different partitions, there nonetheless exist strict inequalities limiting the possible vectors to a non-trivial cone~\cite{CadneyHuberLindenWinter2014}. A consistent generalisation of multipartite entanglement dimensionality can then be given as $d_{\mathrm{GME}}(\rho):=\inf_{\mathcal{D}(\rho)}\max_{|\psi_i\rangle\in\mathcal{D}(\rho)}\min_{\alpha_i}r_{\alpha_i}(|\psi_i\rangle)$.


\textbf{GME classes}.\ While the tensor rank and Schmidt rank vector give further insight into multipartite entanglement structures beyond qubits, there is still a more complex structure hidden beneath. This was first realised in the seminal papers of D{\"u}r, Vidal \& Cirac~\cite{DuerVidalCirac2000} and Ac{\'i}n \textit{et al.}~\cite{AcinBrussLewensteinSanpera2001}, proving that even genuinely multipartite states of three qubits can be inequivalent under LOCC with the famous examples of the Greenberger-Horne-Zeilinger (GHZ) state $\ket{\mathrm{GHZ}}:=\tfrac{1}{\sqrt{2}}(\ket{000}+\ket{111})$. and the W-state $\ket{\mathrm{W}}:=\tfrac{1}{\sqrt{3}}(\ket{001}+\ket{010}+\ket{100})$. This already excludes easy operational measures of entanglement that could be interpreted as asymptotic resource conversions, such as in the bipartite case. In other words, there cannot be a single universal multipartite entangled reference state from which every other state can be created via LOCC (such as the maximally entangled state for bipartite systems), at least not without relaxing the conditions on the set of allowed operations beyond LOCC~\cite{Contreras-TejadaPalazuelosDeVicente2019}. While infinitely many states are needed for such a source set in general~\cite{GourKrausWallach2017, SauerweinWallachGourKraus2018}, in particular, if finite rounds of classical communication are permitted~\cite{SpeeDeVicenteSauerweinKraus2017, DeVicenteSpeeSauerweinKraus2017}, many cases allow finding finite ``maximally entangled sets" of resource states to reach every other state (except for some isolated ``islands") via LOCC~\cite{DeVicenteSpeeKraus2013}. Another option are volume--based approaches, such as the volume of all states reachable via LOCC and the volume of all states from which a state can be reached via LOCC~\cite{SchwaigerSauerweinCuquetDeVicenteKraus2015}. States for which the source volume is zero are extremal resources, whereas the target volume gives a good insight into the general utility of resource states for state transformations. Beyond deterministic transformations, one can also ask when a transformation from a state to another is possible probabilistically. This forms the basis for work in the sub-field of entanglement characterisation using stochastic LOCC (SLOCC), which was first solved for four qubits~\cite{VerstraeteDehaeneDeMoorVerschelde2002} and later for all states that allow for a ``normal form", i.e., which can be filtered to locally maximally mixed states~\cite{VerstraeteDehaeneDeMoor2003}, comprising all states except for a measure-zero subset.


\textbf{Maximal entanglement}.\ While the previous examples show that a universal notion of maximal entanglement cannot exist in the context of LOCC resource theories, one can in principal define states to contain the maximum amount of entanglement if they are maximally entangled across every bipartition. Such states are used in quantum error correction~\cite{Scott2004} and quantum secret sharing~\cite{HelwigCuiLatorreRieraLo2012} and are called \emph{absolutely maximally entangled} (AME) states. It can be shown that for every number $n$ of parties, there is a local dimension $d$ admitting an AME state. However, for $n$ qubits AME states only exist for the $n=2,3,5,6$~\cite{HuberGuehneSiewert2017}.


\textbf{Monogamy of entanglement}.\ Another signature of entanglement in multipartite systems is the phenomenon commonly referred to as \emph{monogamy of entanglement}. The name alludes to the fact that entanglement is not arbitrarily sharable among many parties. To illustrate this point an often invoked example is that of two parties, Alice and Bob, sharing a maximally entangled state $\rho\subtiny{0}{0}{AB}$ such that $E\subtiny{0}{0}{A\!:\!B}(\rho\subtiny{0}{0}{AB})=\log_2(\min[d\subtiny{0}{0}{A},d\subtiny{0}{0}{B}])$. This precludes any further entanglement with a third party. This example, however, is strictly true if and only if $d\subtiny{0}{0}{A}=d\subtiny{0}{0}{B}$, in which case maximal entanglement additionally implies purity of the state $\rho_{AB}$ and thus a tensor product structure with respect to any third party. Quantitatively, monogamy relations are often written in the form
\begin{align}
\label{CKW}
   E\subtiny{0}{0}{A\!:\!BC}(\rho\subtiny{0}{0}{ABC})\leq E\subtiny{0}{0}{A\!:\!B}(\rho\subtiny{0}{0}{AB})+E\subtiny{0}{0}{A\!:\!C}(\rho\subtiny{0}{0}{AC})\,,
\end{align}
but, very recently, monogamy relations have also been recast without inequalities~\cite{GourGuo2018, GuoGour2019}.
The first prominent example valid for three qubits is the Coffman-Kundu-Wootters (CKW) relation~\cite{CoffmanKunduWootters2000}, where the respective entanglement measure is the squared concurrence~\cite{Wootters1998}. This was later generalised to $n$-qubits~\cite{OsborneVerstraete2006}, but proven not to hold for qutrits or higher dimensional systems~\cite{Ou2007}. Moreover, it has been shown that monogamy is a feature only for entanglement measures in a strict sense~\cite{StrelstovAdessoPianiBruss2012} and that monogamy and `faithfulness' (in a geometric sense) are mutually exclusive features of entanglement measures in general dimensions~\cite{LancienDiMartinoHuberPianiAdessoWinter2016}. Meanwhile, additive measures, such as the squashed entanglement~\cite{ChristandlWinter2004}, are monogamous for general dimensions. The inequivalence of the two sides of the inequality~(\ref{CKW}) can in fact be used to quantify and classify multipartite entanglement. For the squared concurrence of three qubits, their difference yields the three-tangle, which is non-zero only for GHZ states and can thus be used to distinguish it from biseparable or W states. A prominent property of the tangle is its invariance not only under local unitaries ($SU(d)$), but also under the complexification of $SU(d)$ to $SL(d)$ to encompass stochastic local operations. This led to the general research line of classifying multipartite entanglement in terms of SLOCC using SL(d)-invariant polynomials~\cite{OsterlohSiewert2005, GourWallach2013}.


\begin{WideBoxes}{\rm Genuine multipartite entanglement in trapped-ion qubits}{GMEions}
Ion trap platforms have been designed primarily for the purpose of fault-tolerant quantum computation~\cite{Bermudez-Mueller2017, BruzewiczChiaveriniMcConnellSage2019} and quantum simulation~\cite{BlattRoos2012}. The main goal is thus to realize a register of individually addressable qubits on which arbitrary quantum gates can be applied. The physical setup usually consists of a linear chain of laser-cooled ions confined by an arrangement of static and oscillating electric fields referred to as a Paul trap. The combination of fields along the principal trap axis and the Coulomb repulsion leads to spatial separation of the individual ions. For each ion, a single qubit is encoded in two long-lived states of the outer valence electron, while additional electronic states are used for shelving and measurements. Local projective measurements on each qubit are performed by detecting spatially resolved, scattered fluorescent light on CCD cameras. For more details on the setup see, e.g.,~\cite{BlattWineland2008}. Depending on the specific choice of atomic species and electronic energy levels, the qubit transitions can either be driven by spatially addressing the ions with laser light (e.g., the setups used in~\cite{Gaebler-Wineland2016, BallanceEtAl2016, JurcevicEtAl2014, Kaufmann-Poschinger2017}), or (near-field~\cite{OspelkausEtAl2011, HartyEtAl2016}, or far-field~\cite{TimoneyEtAl2011, PiltzEtAl2014, WeidtEtAl2016}) frequency addressed microwave signals.

Although the generation of GME is not necessarily the \emph{raison d'$\hat{e}$tre} for many current multi-qubit devices, the controlled generation and detection of GME is often considered as a means to benchmark their functionality~\cite{FriisMartyEtal2018, ZhouZhaoYuanMa2019, WaegellDressel2018}. Consequently, a first generation of ion trap GME experiments has focused on the generation and detection of specific GME states, resulting in the observation of genuine 6-partite GHZ-type entanglement~\cite{Leibfried-Wineland2005}, $8$-qubit W-type GME~\cite{Haeffner-Blatt2005}, with a record of 14-partite GHZ-type GME~\cite{MonzEtAl2011}. Here, GME close to GHZ states can be detected with relatively few measurements (e.g., computational basis measurements plus parity oscillations~\cite{Leibfried-Wineland2005, MonzEtAl2011}) using standard GME witness constructions. Nonetheless, a better characterisation of the produced states and their entanglement structure can be obtained via full state tomography~\cite{Kaufmann-Poschinger2017}, but this approach quickly reaches its practical limits~\cite{Haeffner-Blatt2005}, since the number of measurement settings (here corresponding to global product bases with local dimension $d=2$, see Table~\ref{table:ms}) grows as $3^{N}$ with the number of qubits. Here, matrix product state (MPS) tomography~\cite{CramerPlenio2010, CramerPlenioFlammiaSommaGrossBartlettLandonCardinalPoulinLiu2010, FlammiaGrossBartlettSomma2010} can provide some relief, offering a useful pure state estimate in systems with finite interaction range, e.g., as demonstrated for 14 trapped-ion qubits~\cite{LanyonEtAl2017}, but this is not feasible for 20 qubits~\cite{FriisMartyEtal2018}.

With increasing size of the qubit registers and the desire to certify more complex (multipartite) entanglement structures (e.g., as encountered in quantum simulation~\cite{JurcevicEtAl2014}), it hence becomes necessary to identify simple witnesses based on few measurements. In Ref.~\cite{FriisMartyEtal2018}, such GME witnesses were constructed from fidelities to the closest two-qubit Bell states, averaged over all qubit pairs in groups of $k$ neighbours within a $20$-qubit chain. Intuitively, these witnesses can be understood as a form of monogamy: $2$-qubit entanglement between any pair in a group does not imply GME, but average $2$-qubit entanglement beyond certain thresholds is not compatible with an overall biseparable state. With this approach, genuine tripartite entanglement could be detected simultaneously for every triple of neighbouring qubits in a chain of $20$, making use of measurements in only $3^{3}$ (out of $3^{20}$) global product bases. Using numerical search for $k$-body GME witnesses, the same data could be used to show the development of GME among most neighbouring quadruplets, and some quintuplets.
%
\begin{center}
    \begin{minipage}{0.49\textwidth}
        \begin{center}
            \includegraphics[width=0.99\textwidth]{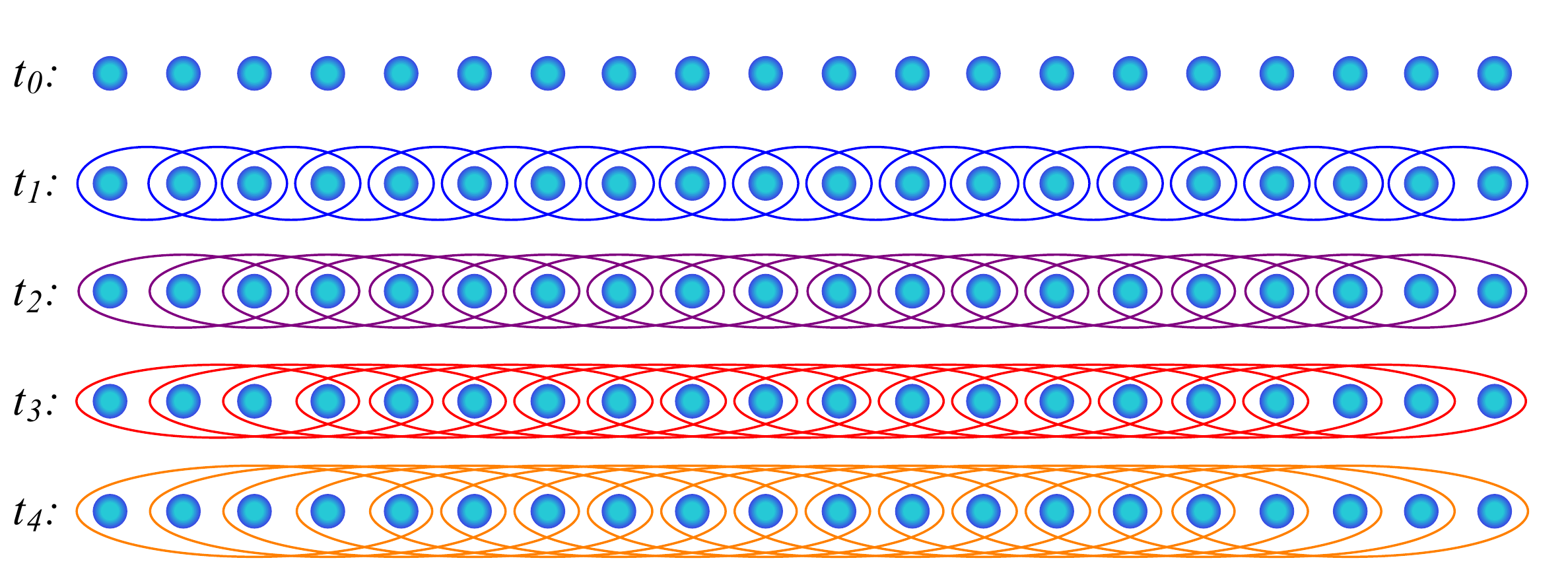}
        \end{center}
    \end{minipage}
    \hspace*{2mm}
    \begin{minipage}{0.48\textwidth}
        \begin{small}
            The illustration shows a simplified schematic of the (genuine multipartite) entanglement structure observed to develop over time under the out-of-equilibrium dynamics of an Ising-type Hamiltonianin~\cite{FriisMartyEtal2018}. There, a chain of $20$ initially separable qubits evolves into states with bipartite entanglement between all neighbours, and consecutively GME between neighbouring groups of 3, 4 and 5 qubits over the course of several independently measured time steps $t_{i}$.
        \end{small}
    \end{minipage}
\end{center}
\end{WideBoxes}


\textbf{PPT mixers}.\ Analogous to the bipartite case, the convex structure of (partial)-separability permits the construction of multipartite entanglement witnesses. However, the additional challenge of the potentially different partitions of density matrix decomposition elements prevents the applicability of many techniques for bipartite witnesses in multipartite systems. In particular, positive maps and their resulting witnesses are inherently connected to bipartite structures. Nonetheless they can be harnessed as constraints for positive-semidefinite programming (SDP). This follows from the simple observation that a state that is decomposable into bi-product states is,  for instance, also decomposable into PPT states. This insight has led to the concept of PPT mixers~\cite{JungnitschMoroderGuehne2011a, JungnitschMoroderGuehne2011b}, yielding effective numerical tools for low dimensions. At the same time, this connection can be used to effectively 'lift' bipartite witnesses for multipartite usage~\cite{HuberSengupta2014, LancienGuehneSenguptaHuber2015}, and to obtain generalisations to maps which are positive on biseparable states~\cite{ClivazHuberLamiMurta2017}.


\textbf{GME witnesses}.\ A canonical form of GME-witnesses can be obtained by harnessing the different Schmidt decompositions across bipartitions. For instance, for a pure target state $\ket{\psi\subtiny{0}{0}{\mathrm{T}}}$, computing all marginal eigenvalues allows defining a witness~\cite{Bourennane-Sanpera2004} of the form $W_{\mathrm{GME}}:=\max_{\alpha_i}||\rho_{\alpha_i}||_{\infty}\mathbbm{1}-\ket{\psi\subtiny{0}{0}{\mathrm{T}}}\!\!\bra{\psi\subtiny{0}{0}{\mathrm{T}}}$. Apart from this generically applicable method, most available GME witnesses are tailored towards detecting specific 
states, such as graph states~\cite{HeinDuerEisertRaussendorfVanDenNestBriegel2006} or stabilizer states~\cite{TothGuehne2005b, AudenaertPlenio2005, SmithLeung2006}, Dicke states~\cite{BergmannGuehne2013}, or generally symmetric states~\cite{TothGuehne2009}.

Leaving the regime of linear operators and moving on to non-linear functions of density matrix elements, more powerful certification techniques exist. In~\cite{GuehneSeevinck2010} and~\cite{HuberMintertGabrielHiesmayr2010} non-linear inequalities for detecting multipartite entanglement in GHZ and W like states were introduced, which were proven to be strictly more powerful than the canonical form introduced above. Moreover, these non-linear inequalities were later shown to provide lower bounds on a particular measure of genuine multipartite entanglement, the \emph{GME}-concurrence~\cite{MaChenChenSpenglerGabrielHuber2011}. In fact, one can leverage the previously mentioned SDP techniques to numerically evaluate multiple suitable convex-roof-based entanglement measures~\cite{TothMoroderGuehne2015}. In a separate approach, \emph{separability eigenvalues} were introduced as a means to construct multipartite entanglement witnesses~\cite{SperlingVogel2013}.


\textbf{Entanglement and spin-squeezing}.\ A prominent example for a many-body system that is quite well understood from both the theoretical and experimental perspective is an ensemble of $N\gg 1$ (``pseudo")-spins manipulated (and measured) collectively in a localized trap. To detect entanglement, spin-squeezing criteria for entanglement have been derived. These are based on an analogy with bosonic quadratures and are connected with uncertainty relations of collective spin components.
Most famously, a necessary condition for all fully separable states of $N$ particles with spin-$1/2$ reads $\xi_{\rm S}^2:=N \tfrac{(\Delta J_z)^2}{\langle J_x\rangle^2+\langle J_y\rangle^2} \geq 1$, which also directly connects entanglement with enhanced sensitivity in Ramsey spectroscopy with totally polarised ensembles of atoms~\cite{KitagawaUeda1993, WinelandBollingerItanoHeinzen1994, SoerensenDuanCiracZoller2001, SoerensenMoelmer2001}. Here, $(\Delta J_z)^2$ is the smallest variance in a direction orthogonal to the polarisation, say, $|\langle J_y \rangle|\approx N/2$ and a spin-squeezed state (SSS) is obtained when $\xi_{\rm S}^2<1$, where the boundary value defines the coherent spin states (CSS). After the first pioneering works, the concept of spin-squeezing has been generalised in several directions~\cite{Ma2011Quantum, 2016arXiv160901609P}.

As a generalization, a full set of spin-squeezing inequalities, which have the geometrical shape of a closed convex polytope and define a more general spin-squeezing quantifier, have been derived for spin-$\frac{1}{2}$ ensembles~\cite{tothPRL07,tothPRA09} and later generalised to all higher spin-$j$ ensembles and also to $su(d)$ observables different from angular momentum components~\cite{vitagliano11, vitagliano14}. Thus, witnessing entanglement via the concept of squeezing of the collective spin of an ensemble can be convenient because this notion is captured by a simple polytope in the space of collective spin variances. A similar simple structure remains even for device-independent certification of entanglement based on collective measurements~\cite{Tura-Acin2014}.

\emph{Entanglement depth} is typically used as a quantifier of entanglement in spin-squeezed states, which can also be witnessed with spin-squeezing parameters by making use of the Legendre transform method~\cite{SoerensenMoelmer2001, vitagliano16, VitaglianoPlanar}. The general picture is that one can find a hierarchy of bounds on some collective quantities which depend on the entanglement depth, like $(\Delta J_z)^2 \geq Nj F_{J}\left(\frac{\langle J_y \rangle}{Nj}\right)$,
where $F_J$ is a certain convex function which can be obtained through Legendre transforms. A state with the property that the variance on the left-hand side is below the value of the right-hand side for a certain $F_J$ is detected with a depth of entanglement of at least $k=J/j$, where $j$ is the spin quantum number of the individual particles. Entanglement depth criteria similar to the above have also been derived for different target states, like Dicke states~\cite{Lucke2014Detecting, vitagliano16} and planar quantum squeezed states~\cite{HePRA2011, VitaglianoPlanar}, and also based on other quantities, like the QFI~\cite{Pezze2009Entanglement, HyllusLaskowskiKrischek12, TothEntanglementMetrologyFisher12, GessnerPezzeSmerzi16, GessnerPezzeSmerzi17}.


\textbf{Entanglement in optical lattices}.\ Beyond clouds of atoms or BECs in single localized traps, a current challenge is to demonstrate and exploit multipartite entanglement in spatially extended systems, such as optical lattices. Here, as well as for localized traps, the most common measurement consists of releasing the gas from the trap (i.e., the lattice potential) and imaging the expanding gas, inferring the momentum distribution of the original system of particles. Besides spin-squeezing methods that could also be employed in these systems, criteria to detect entanglement in optical lattices have been proposed based on quantities obtained from density measurements after a certain time of flight~\cite{VollbrechtCirac2007, CramerPlenioWunderlich11}. Furthermore, some collective quantities with thermodynamical significance, like energy~\cite{DowlingDohertyBartlett2004,tothpra05}, or susceptivities~\cite{Wiesniak05, brukner06, HaukeHeylTagliacozzoZoller16} (e.g., to external magnetic fields) could be used for entanglement detection in such extended systems. These quantities can be extracted from, e.g., the structure factors coming from neutron scattering cross sections~\cite{brukner06, CramerPlenioWunderlich11, CramerEtAl2013, marty14, MartyVitagliano}. Some of these methods have been employed for a first experimental demonstration (and quantification) of entanglement in a bosonic optical lattice~\cite{CramerEtAl2013}, while other recent experiments~\cite{FukuharaHildZeiherSchaussBlochEndresGross2015, DaiEtAl2015, IslamMaPreissTaiLukinRispoliGreiner2015} demonstrate entanglement between two spins in a (super)lattice.


\begin{WideBoxes}{\rm Cold-atom entanglement in the lab}{coldatombox2}
Experimentally, entanglement through spin-squeezing has been demonstrated extensively in atomic ensembles (see, e.g., the reviews~\cite{2016arXiv160901609P, Ma2011Quantum}). To summarise shortly, dynamics used can be split into two main groups: a) atom-atom interactions in Bose-Einstein Condensates, and b) light-atom interactions in ensembles of room temperature or cold gases.
In such systems, state tomography is usually performed in a collective Bloch sphere of three orthogonal collective spin directions. The mean spin polarisation $\langle \vec J \rangle=(\langle J_x \rangle,\langle J_y \rangle,\langle J_z \rangle)$ can be depicted as a vector, together with variances $(\Delta J_k)^2$ as uncertainty regions around it~\cite{2016arXiv160901609P, Ma2011Quantum}. Note that other collective $su(2j+1)$ operators, and thus correspondingly different collective Bloch spheres, have also been recently considered in spinor ensembles~\cite{HamleyGervingHoangBookjansChapman2012, KunkelEtAl2018}.

A prominent example of dynamics that produce (in terms of spectroscopic gain below the Standard Quantum Limit, up to some $dB$s of) spin-squeezing via atom-atom interaction is the \textit{one-axis twisting} dynamics, $H_{\rm OT}\propto J_x^2$ employed in several experiments with BECs~\cite{Orzel2386, Esteve2008Squeezing, Riedel2010Atom-chip-based, Gross2010Nonlinear, OckeloenSchmiedRiedelTreutlein2013, BerradaVanFrankBueckerSchummSchaffSchmiedmayer2013, MuesselStrobelLinnemannHumeOberthaler2014, SchmiedBancalAllardFadelScaraniTreutleinSangouard2016}. For the second group, a widely used method is the production of (similarly, below the SQL of several, up to the tens of $dB$s of) spin-squeezing via \textit{quantum non-demolition} (QND) measurement and feedback, which consists of sending pulses of light through the ensemble of atoms and engineering the interaction $H_{\rm QND}\propto S_z J_z$ which rotates the light polarisation and conserves $J_z$. This technique has been used in cold as well as room temperature atomic ensembles~\cite{KuzmichMandelBigelow2000, AppelWindpassingerOblakHoffKjaergaardPolzik2009, TakanoFuyamaNamikiTakahashi2009, TakanoTanakaNamikiTakahashi2010, SewellNapolitanoBehboodColangeloMartinCiuranaMitchell2014, InoueTanakaNamikiSagawaTakahashi2013, SewellKoschorreckNapolitanoDubostBehboodMitchell2012, BaoEtAl2018}, also with an additional coupling to an optical-cavity, which enhances the optical depth of the ensemble~\cite{SchleierSmithLerouxVuletic2010, LerouxSchleierSmithMonikaVuletic2010, ChenBohnetSankarDaiThompson2011, ZhangMcConnellCukvSchleierSmithLerouxVuletic2012, BohnetCoxNorciaWeinerChenThompson2014, HostenEngelsenKrishnakumarKasevich2016, CoxGreveWeinerThompson2016}. Notably, entanglement (with a depth up to few thousands) has also been achieved via other regimes of light-mediated atomic interactions~\cite{FernholzKrauterJensenShersonSoerensenPolzik2008, HaldSoerensenSchoriPolzik1999, KuzmichMoelmerPolzik1997, McConnellZhangHuCukVuletic2015} and QND measurements have also been used to entangle two macroscopic room temperature vapour cells~\cite{JulsgaardKozhekinPolzik2001, JulsgaardShersonCiracFiurasekPolzik2004}.

Recently, generalised spin-squeezed states, like \textit{singlet states}~\cite{BehboodMartinCiuranaColangeloNapolitanoTothSewellMitchell2014} or \textit{planar squeezed states}~\cite{ColangeloCiuranaBianchetSewellMitchell2017, VitaglianoPlanar}, have been investigated in experiments with atomic ensembles and have been proposed for application in quantum metrology. In particular, let us emphasise \textit{Dicke states} in this context, which are attracting increasing attention and are produced (up to the ten of $dB$ of squeezing with respect to a SQL) in experiments with Bose-Einstein condensates~\cite{LueckeEtAl2011, HamleyGervingHoangBookjansChapman2012, Lucke2014Detecting, PeiseEtAl2015}, with atomic \textit{spin-mixing dynamics} resembling parametric down-conversion of photons to some extent. A depth of entanglement of several hundreds has been inferred with collective measurements also for  these generalised spin-squeezed states~\cite{Lucke2014Detecting, HoangEtAl2016, LuoEtAl2017, EngelsenKrishnakumarHostenKasevich2017, VitaglianoPlanar}. Finally, current experimental efforts have been oriented towards demonstrating entanglement between spatially separated parts of BECs (still in localized traps)~\cite{LangeEtAl2018, KunkelEtAl2018, FadelZiboldDecampsTreutlein2018}.
\begin{center}
    \begin{minipage}{0.31\textwidth}
        \begin{center}
            \includegraphics[width=0.99\textwidth]{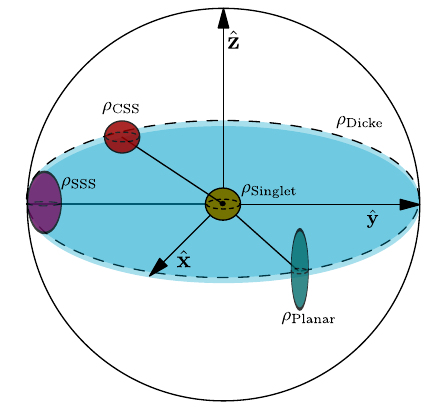}
        \end{center}
    \end{minipage}
    \hspace*{2mm}
    \begin{minipage}{0.58\textwidth}
        \begin{small}
            Generalised spin-squeezed states in the collective Bloch sphere. In the illustration of generalised spin-squeezed states in the collective Bloch sphere, states are represented as vectors (for the global spin length $\langle \vec J \rangle$) with uncertainty regions around it. These regions also take into account classical (usually Poissonian) noise: (i) $\rho_{\rm CSS}$ is a completely polarised, mixed state $|\langle \vec{J} \rangle |\simeq O(N)$ close to a CSS that has three variances of the order of $(\Delta J_k) \simeq O(\sqrt N)$, (ii) $\rho_{\rm SSS}$ is completely polarised and has a single squeezed variance in a direction orthogonal to its polarisation, (iii) $\rho_{\rm Planar}$, a planar squeezed state, almost completely polarised with two squeezed variances, (iv) $\rho_{\rm Singlet}$, a macroscopic singlet state, with all three variances squeezed, and (v) the unpolarised Dicke state $\rho_{\rm Dicke}$, with a tiny uncertainty $(\Delta J_z) \simeq 0$ and large $(\Delta J_x)=(\Delta J_y)\simeq O(N)$.
        \end{small}
    \end{minipage}
\end{center}
\end{WideBoxes}


\vspace*{-2mm}
\section*{Conclusion and Outlook}
\vspace*{-2mm}

As this review shows, the certification of entanglement is a highly active field with current challenges ranging across a diverse set of topics. From the humble beginnings in the foundation of quantum mechanics multiple sub-fields have emerged that could not possibly be explained in a single review. We have thus focused only on the theoretical methods and experimental platforms for efficient certification and quantification in contemporary quantum technologies with a particular focus on the high-dimensional and multipartite case.

For the sake of brevity, we have mainly discussed the case of well-characterised measurement devices and system Hamiltonians. It is indeed possible to transcend this paradigm and obtain robust entanglement certification techniques that do not require a detailed physical understanding of the measurement procedure that is used or the system that is investigated. These \emph{device-independent} certification techniques currently require a larger amount of resources and suffer from poor robustness to experimental noise. As the proficiency in handling quantum technologies increases, the logical next step is to move step-by-step towards more device-independent certification techniques, increasing the security in both quantum communication itself and in our confidence in the correct functionality of quantum devices.

Finally, while the use of bipartite high-dimensional entanglement is well established, the unfathomable complexity of multipartite quantum correlations has so far only found few applications in many-party protocols, and for some applications they may not be useful at all (such as, e.g., universal quantum computation~\cite{GrossFlammiaEisert2009}). Finding further compelling quantum information protocols would motivate a deeper look into the structure of multipartite entanglement and guide theoretical as well as experimental efforts towards the preparation, manipulation, and certification of novel many-body quantum states.\\
\vspace*{-3mm}


\textbf{Acknowledgements}.
We acknowledge support from the Austrian Science Fund (FWF) through the START project Y879-N27, and the joint Czech-Austrian project MultiQUEST (I3053-N27 and GF17-33780L). N.F. acknowledges support from the Austrian Science Fund (FWF) through the project P 31339-N27. M.M. acknowledges support from the QuantERA ERA-NET Co-fund (FWF Project I3773-N36) and the UK Engineering and Physical Sciences Research Council (EPSRC) (EP/P024114/1). G.V. acknowledges support from the Austrian Science Fund (FWF) through the Lise-Meitner project M 2462-N27.


\bibliography{samplearxiv}


\end{document}